\documentclass[%
 reprint,
superscriptaddress,
 amsmath,amssymb,
 aps, 
floatfix,
]{revtex4-2}

\usepackage{mathtools}
\usepackage[english]{babel}
\usepackage{graphicx}

\usepackage[colorlinks=true, allcolors=blue]{hyperref}
\usepackage[capitalise]{cleveref}

\usepackage[dvipsnames]{xcolor}
\newcommand{\edit}[1]{#1}

\newcommand{\boldsf}[1]{\textsf{\textbf{#1}}}

\begin{document}



\title{\edit{Intrinsic statistical separation of subpopulations in heterogeneous collective motion via dimensionality reduction}}

\affiliation{
 Mathematical, Computational, and Systems Biology Graduate Program,}
\affiliation{Department of Mathematics,}
\affiliation{University of California, Irvine.}%

\author{Pei Tan}
\email{peit3@uci.edu}
\affiliation{
 Mathematical, Computational, and Systems Biology Graduate Program,}
\affiliation{University of California, Irvine.}%
\author{Christopher E. Miles}
 \email{chris.miles@uci.edu}
\affiliation{Department of Mathematics,}
\affiliation{University of California, Irvine.}%

\date{\today}

\begin{abstract}
Collective motion of locally interacting agents is found ubiquitously throughout nature. The inability to probe individuals has driven longstanding interest in the development of methods for inferring the underlying interactions. In the context of heterogeneous collectives, where the population consists of individuals driven by different interactions, existing approaches require some knowledge about the heterogeneities or underlying interactions. Here, we investigate the feasibility of identifying the identities in a heterogeneous collective without such prior knowledge. We numerically explore the behavior of a heterogeneous Vicsek model and find sufficiently long trajectories \edit{intrinsically cluster in a PCA-based dimensionally reduced model-agnostic description of the data.} We identify how heterogeneities in each parameter in the model  (interaction radius, noise, population proportions) dictate this clustering. Finally, we show the generality of this phenomenon by finding similar behavior in a heterogeneous D'Orsogna model. Altogether, our results \edit{establish and quantify the intrinsic model-agnostic statistical disentanglement of identities in heterogeneous collectives.}\end{abstract}

\maketitle

\section{Introduction}

Systems of locally interacting agents that display spatiotemporal collective behaviors beyond the capabilities of individuals are found ubiquitously throughout the physical world at a range of scales~\cite{vicsek2012collective,deutsch2020MultiscaleAnalysisModelling}. Notable examples include fish schooling~\cite{fish2004,jhawar2020NoiseinducedSchoolingFish}, birds flocking~\cite{bialek2012StatisticalMechanicsNatural,bird2019}, insect~\cite{bernoff2020AgentbasedContinuousModels,weinburd2021AnisotropicInteractionMotion} and bacterial swarming~\cite{bacterial2010,bacterial2012}, human crowds~\cite{rio2018LocalInteractionsUnderlying}, cell migration~\cite{vicsek2014,migration2020}, and other subcellular processes~\cite{schaller2010polar,miles2022MechanicalTorquePromotes}.

Most attention has been paid towards investigating homogeneous collectives, where all agents evolve and interact via the same dynamics. However, real collectives are richly heterogeneous~\cite{jolles2020role,ariel2022variability}. Such heterogeneities arise from bacterial length differences~\cite{peled2021HeterogeneousBacterialSwarms}; mixed-species collectives~\cite{ward2018cohesion}; leader-follower behaviors in animals~\cite{herbert-read2013RoleIndividualityCollective, collignon2019collective,mizumoto2021CoordinationMovementComplementary,gomez-nava2022IntermittentCollectiveMotion} or cell migration~\cite{schumacher2017SemblanceHeterogeneityCollective,fu2018SpatialSelforganizationResolves,kwon2019stochastic,qin2021roles}; lane formation in human crowds~\cite{zhang2019pedestrian}. The collective motion of heterogeneous systems has consequently been investigated extensively and found to be even richer than that of the homogeneous variety~\cite{phase2015,copenhagen2016SelforganizedSortingLimits,del2018importance,hoell2019MultispeciesDynamicalDensity,netzer2019HeterogeneousPopulationsNetwork,heter2022}.

Alongside the studies of the emergent behavior of collectives, a parallel thread of investigations has developed and applied methods for the inverse problem of deducing the underlying interactions from trajectories~\cite{lukeman2010InferringIndividualRules,mann2011BayesianInferenceIdentifying,herbert-read2011InferringRulesInteraction, katz2011InferringStructureDynamics,gautrais2012DecipheringInteractionsMoving,lord2016InferenceCausalInformation,torney2018InferringRulesSocial,lu2019nonparametric,d2019,basak2020InferringDomainInteractions,lachance2022LearningRulesCollective,nabeel2023DatadrivenDiscoveryStochastic}. This quest is of natural scientific interest due to the ability to observe only the correlated trajectories of the interactive collective, making disentangling individual interactions inherently challenging, especially with heterogeneities \cite{schumacher2017SemblanceHeterogeneityCollective}. Recent advances have broken ground on the ability to infer interactions in heterogeneous collectives using clever and sophisticated approaches. However, these approaches, while powerful and elegant, seemingly share a unifying feature of requiring knowledge of the collective or its heterogeneities. For instance, methods that provide flexible non-parametric tests of heterogeneities~\cite{schaerf2021StatisticalMethodIdentifying}, or the ability to infer the interactions~\cite{luLearningInteractionKernels} in heterogeneous collectives, both require knowledge of the particle identities \textit{a priori}. The work in~\cite{Messenger2022} addresses this with a mixture model fit alongside sparse identification of the interactions. While able to identify the identities, the success of this method hinges on the ability to correctly specify a library of underlying interactions. Other methods for detecting heterogeneities work well but are limited to specific contexts such as the detection of dissenting directions among neighbors~\cite{nabeel2022DisentanglingIntrinsicMotion} or only leader-follower interactions \cite{butail2016model,mwaffo2017analysis}. In this work, we seek to address whether particle identities can be detected in heterogeneous collectives with no prior information about the collective or the structure of the heterogeneities. 

\begin{figure}[b]
    \centering
    \includegraphics[width=\linewidth]{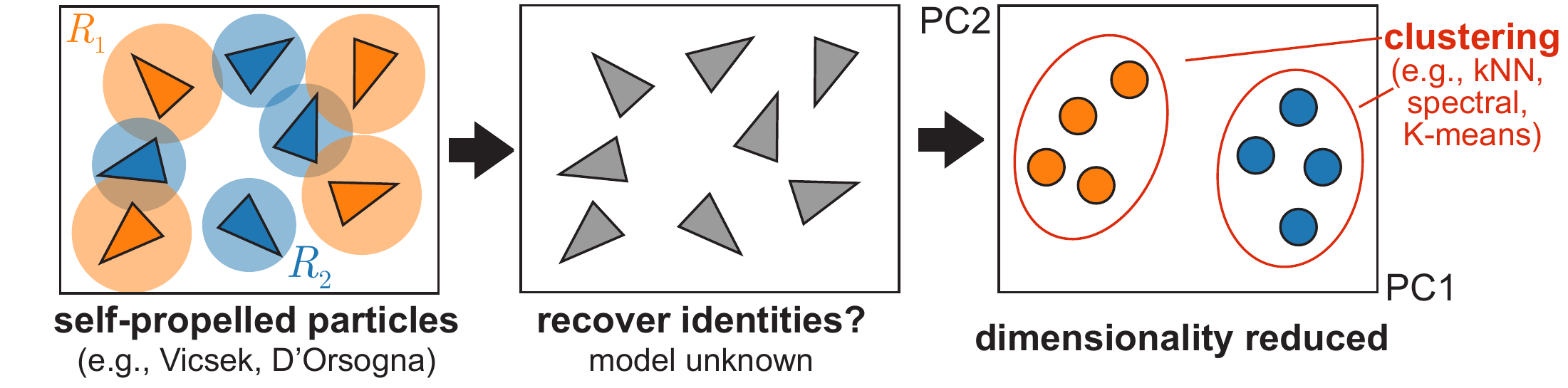}
    \caption{\label{fig:flow} \edit{\textbf{Schematic of the investigation.}
   The central question of this work is whether (and how) the identities of particles in heterogeneous collectives (e.g., a Vicsek model with two different interaction radii for each subpopulation $R_1, R_2$) can be recovered from trajectory data with no model information. We find that dimensionality reduction via PCA (principal component analysis) yields distinct clustering of the subtypes over sufficiently long timescales characterized in our work.}}
   
    \end{figure}

To study disentangling heterogeneities in collectives, we investigate a heterogeneous variant of the classical Vicsek model \cite{Vicsek1995}. This model is renowned as the textbook minimal example of a collective motion with rich behavior~\cite{physics2016,vicsek2000}. Consequently, many variants have been considered \cite{chate2008modeling}, including those with heterogeneities \cite{miguel2018effects,chatterjee2023flocking} such as the ones we propose here. We first consider a setup with two populations of Vicsek particles with different parameters (interaction radii, noise magnitude, velocity), but still interacting as a single collective. After performing dimensionality reduction on the trajectories, we find that in this latent space, the trajectories cluster into their identities for sufficiently long observations. In this work, we quantify the parameter-dependent timescale required for accurate clustering through numerical simulation. Next, we show that this clustering phenomenon persists in a heterogeneous Vicsek model with more than two species. Lastly, to establish that this is truly a model-free phenomenon, we consider a heterogeneous D'Orsogna model \cite{dorsogna2006SelfPropelledParticlesSoftCore} and find similar clustering behavior. Altogether, our results are summarized in \cref{fig:flow} and establish the ability to cluster heterogeneous collectives in a model-free manner with no prior knowledge of the underlying model or heterogeneities.

\section{Setup}

\subsection{Classical Vicsek}

The classical Vicsek model describes the evolution of $N$ self-propelled particles moving in 2-dimensional space at a constant speed $\nu$ and with fluctuating direction. The direction of each particle is governed by two factors: noise, and local interactions with neighbors. Specifically, each particle averages the orientations over all neighbors within a specified radius, $R$. In symbols, $\theta_{i,t}$, the orientation of particle $i$ at frame $t$, evolves as 
\begin{equation}
\label{eq:vicsek_vel}
\theta_{i,t+1} = \langle \theta_{j,t}(t) \rangle_{\|\pmb{x}_{i,t}-\pmb{x}_{j,t}\|<R} + \eta.
\end{equation}
The particle positions are updated with these orientations
\begin{equation}
\label{eq:vicsek_pos}
\pmb{x}_{i,t+1} =  \pmb{x}_{i,t} + \nu \Delta t  \begin{pmatrix} \cos(\theta_{i,t}) \\ \sin(\theta_{i,t})  \end{pmatrix} 
\end{equation}
 The noise $\eta$ is chosen from a uniform distribution governed by a scalar magnitude $0 \leq  \sigma \leq 1$, such that $\eta \sim \mathrm{unif}(-\sigma \pi, \sigma \pi)$. The particles are constrained to an $L \times L$ periodic box, where distances are computed in a manner that respects the periodicity of the domain. For systems with large $N$, naive $\mathcal{O}(N^2)$ comparisons are prohibitive. We instead employ a standard KD-tree \cite{brown2017review}  $\mathcal{O}(n\log n)$ implementation for computational scalability. \edit{Particles are initialized with uniformly random orientation and position within the box. For all simulations, unless noted otherwise, $t=1000$ steps are taken for equilibration and then discarded for analysis. This choice is discussed further in the text in  \cref{subsect:transient}.}

\subsection{Heterogeneous Vicsek and clustering pipeline}

We consider a variant on the classical Vicsek model with $M \geq 2$ subpopulations. Specifically, denote $\pmb{\phi}=(\nu, \sigma, R)$ as the parameters governing the motion of a particle in the classical Vicsek model. In the heterogeneous collective, particles belonging to subpopulation $j$ evolve via the parameter set $\pmb{\phi}_j=(\nu_j, \sigma_j,R_j)$. Particles interact regardless of their membership in a subpopulation. In total, the collective consists of $N$ particles that can be decomposed into their group membership $N=\sum_{j=1}^{M} N_j$, where $N_j$ denotes the number of particles in subpopulation $j$. This model has been considered in previous studies and is a more general case of some leader-follower models.

The heterogeneous Vicsek model is straightforward to simulate and generate trajectories for testing. However, performing the cluster analysis on the resulting trajectories in an unsupervised model-agnostic manner does not seem to have a \edit{clearly outlined path in the existing literature.} 

\edit{
The first design decision we must make is the input data to the procedure. We assume that only positional information is available, and the particle identity is known frame-to-frame, allowing for the formation of trajectories. To reduce each trajectory to a scalar quantity, we consider  $\theta_i(t)$, the orientations. While it may not be possible to directly access these for experimental observations, the orientations can be estimated by the frame-to-frame displacement e.g., $\hat{\theta}_{i,t}= \texttt{atan2}(\pmb{x}^y_{i,t+1} - \pmb{x}^y_{i,t},\pmb{x}^x_{i,t+1} - \pmb{x}^x_{i,t})$, where $\pmb{x}^{x,y}_{i,t}$ correspond to the $x,y$ component of the positions. Naive dimensionality reduction does not preserve the structure of angular data \cite{sargsyan2012GeoPCANewTool}, so we transform $\tau_{i,t} := \tan\theta_{i,t}$. Alternatively, we tested $\tilde{\tau}_{i,t} := [\cos\theta_{i,t},\sin\theta_{i,t}]$, which doubles the trajectory length but may be more generalizable to 3D data, and found no difference in our results. In summary, for $t$ observations of a collective with  $N$ particles, we consider our data to be the $N \times t$ matrix 
\begin{equation}\label{eq:datamat}
    X_t = \tan(\Theta_t) = \begin{bmatrix} 
    \tan\theta_{1,0} & \tan\theta_{1,1} & \cdots & \tan\theta_{1,t}\\
    \tan\theta_{2,0} & \tan\theta_{2,1} & \cdots & \tan\theta_{2,t}\\   
    \vdots & \vdots & \ddots & \vdots \\
       \tan\theta_{n,0} & \tan\theta_{n,1} & \cdots & \tan\theta_{n,t}
    \end{bmatrix}.
\end{equation}}

Equipped with this data, there are two notable branches of approaches for unsupervised clustering time series \cite{aghabozorgi2015time}. One can assign and cluster based on an appropriate metric between trajectories, such as Euclidean distance or dynamic time warping\cite{jeong2011weighted}. However, the choice of such a metric for collective motion data is not obvious to the authors. Therefore, we consider the second main avenue for clustering time series: dimensionality reduction. A zoo of possible linear and nonlinear approaches for dimensionality reduction of time series exists. We opt for  \edit{a pragmatically simple approach of principal component analysis (PCA). While classical, it is worthwhile to note that PCA can outperform nonlinear dimensionality reductions in certain contexts~\cite{zhou2022PCAOutperformsPopular} and has interpretability as linear transformations of the original data. There may be more complex dimensionality reduction procedures that better separate the data, but PCA would nonetheless always be the benchmark to compare the performance with and therefore serve as the basis of the remainder of this work.}

\edit{We briefly review PCA for self-containment of our approach's description. Further details can be found in \cite{pca2022} For the data matrix $X_t$ in \eqref{eq:datamat}, PCA corresponds to a $t \times t$ weight matrix $W_t$, whose columns are the eigenvectors of $X_t^\mathsf{T}X_t$, from which a component matrix $T_t$ can be computed by $T_t = X_t W_t$. Each column of $T_t$ is scaled to have unit variance and zero mean. This construction corresponds to a linear change of basis to orthogonal directions that maximize variances within the data. In practice for dimensionality reduction, only the first $L$ columns of $T_t$ are considered, defining a linear transformation of each row of the data $X_t$ (a particle trajectory) into an $L$ dimensional vector of scores. Throughout the remainder of this work, we consider $L=2$ due to the ability to visualize the scores. However, we found little performance dropoff or gain for larger or smaller values of $L$, including $L=1$ for the heterogeneous Vicsek model.}

\edit{While PCA is notably useful in transforming the data to a more easily clusterable description, it is not itself, a clustering technique. Therefore, we must finally choose some approach for procedurally identifying clusters. In practice, we considered alternatives (K-nearest neighbors~\cite{zhang2017efficient}, spectral clustering~\cite{spec2007}) but find this choice matters very little due to the intrinsic behavior of separation between the two particle populations in PCA space.  Unless otherwise noted, all clustering in the remainder of the text is done using K-means, which assigns $C$ cluster identities $\mathcal{S}=\{S_1,\ldots,S_C\}$ based on the optimization of the total distance away from centroids within each cluster, $\operatorname{arg\,min}_\mathcal{S}  \sum_{i=1}^{C} \sum_{\mathbf x \in S_i} \left\| \mathbf x - \boldsymbol\mu_i \right\|^2$. Here, $\mathbf{x}$ are $L$-dimensional vectors of PC scores for each trajectory, and $\boldsymbol{\mu}_i$ are the centroids (means) computed from the cluster assignments. This optimization is done using \texttt{scikit-learn}'s standard \texttt{KMeans} function with the known number of clusters specified. }

\section{Results}

\subsection{Two subpopulation heterogeneous Vicsek models cluster over sufficiently long times.}

\begin{figure}[htb]
    \centering
    \includegraphics[width=.99\linewidth]{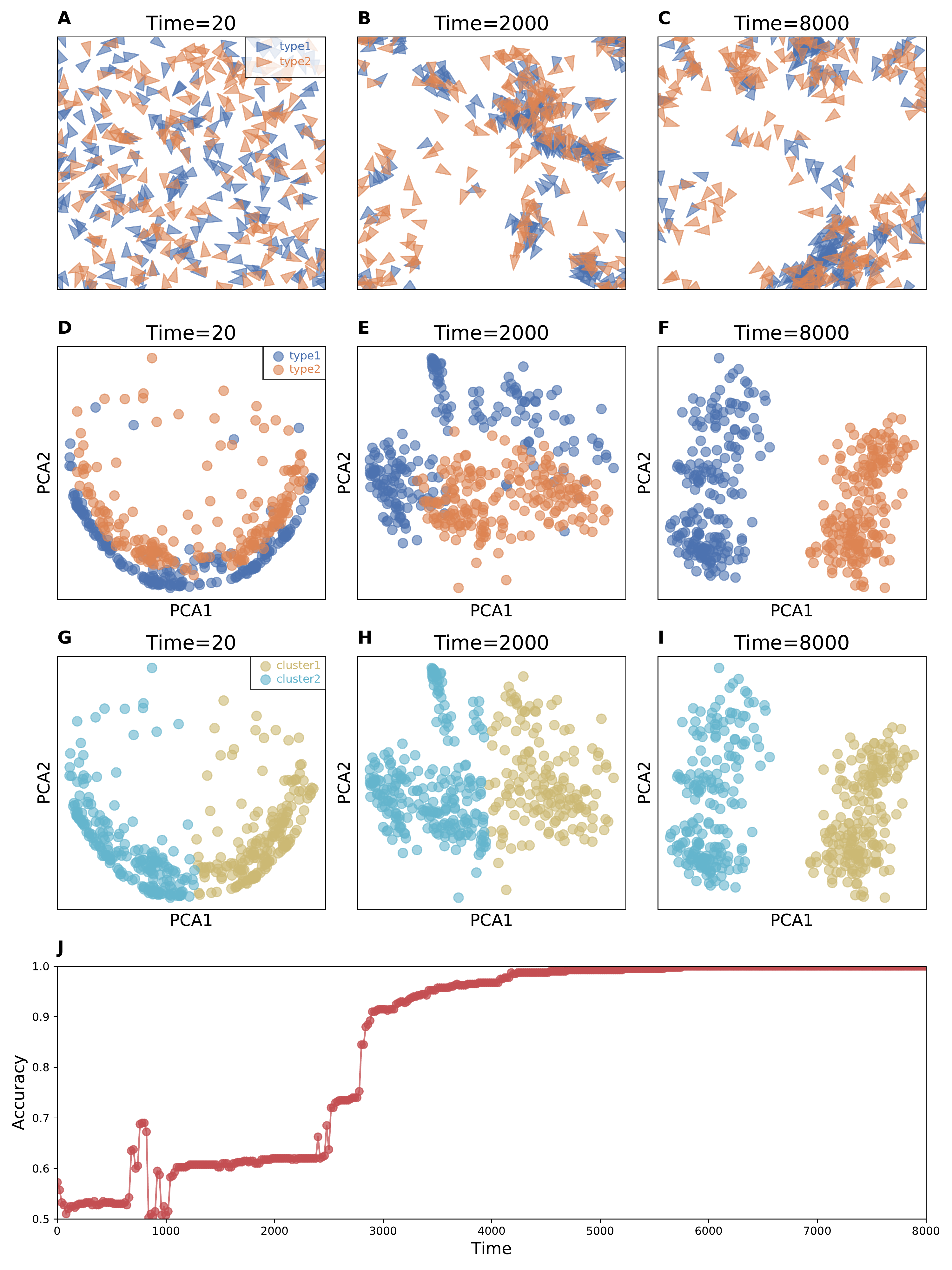}
    \caption{\label{fig:vicsek_double} \textbf{Two subtype Vicsek model simulation and clustering.} 
    \boldsf{ABC}: snapshots of the particle positions in a heterogeneous Vicsek simulation with two types of particles and display no apparent pattern. \boldsf{DEF}: The first two principal component scores for each trajectory, colored by particle type. \boldsf{GHI}: Results of K-means clustering on PC scores. \boldsf{J}: Clustering accuracy approaches 100\% as the trajectories become longer. The two populations differ only in their noise magnitude $\sigma_1 = 0.1, \sigma_2=0.3$ and otherwise $\nu=0.01, R=.05$ with $L=1,\Delta t = 1$, and particle counts$ N_1=200,N_2=200$.}
    \end{figure}

We first demonstrate the dimensionality-reduction-based clustering on a  setup with two subpopulations of particles that differ only in one attribute. Specifically, we take two types of particles, $N_1=200,$$ N_2=200$ with $\pmb{\phi}_1=(\nu_1, \sigma_1,R_1) $$ =(0.01, 0.1, 0.05)$ and $\pmb{\phi}_2$$=(\nu_2, \sigma_2,R_2) = (0.01, 0.3, 0.05)$. That is, the two particles differ only in their magnitude of noise. Other simulation parameters are set to $L=1,  \Delta t = 1$.  The results of the simulation over increasingly long times can be seen in \cref{fig:vicsek_double}. 

In panels \boldsf{ABC} of \cref{fig:vicsek_double}, the snapshots of particle positions show that the collective evolves in a manner that integrates both subpopulations with no apparent pattern. The first two principal component scores of each trajectory are shown in panels \boldsf{DEF}. At early times, the scores are not separable by eye. After some time passes, the scores seem to begin to separate but not to a degree that can be fully disentangled. Finally, at long times, the PC scores of the trajectories corresponding to different types separate into two distinct clusters. Panels \boldsf{GHI} show the result of running K-means clustering on the PC scores. Initially, the clustering is inaccurate \edit{(around 50\%, as expected, by random assignments of two categories)} but progressively gains accuracy until eventually stabilizing at 100\% as more data is accumulated on the trajectory (panel \boldsf{J}). 

\subsection{Time to accurately cluster is dependent on which parameters are heterogeneous.}

\begin{figure}[htb]
    \centering
    \includegraphics[width=0.99\linewidth]{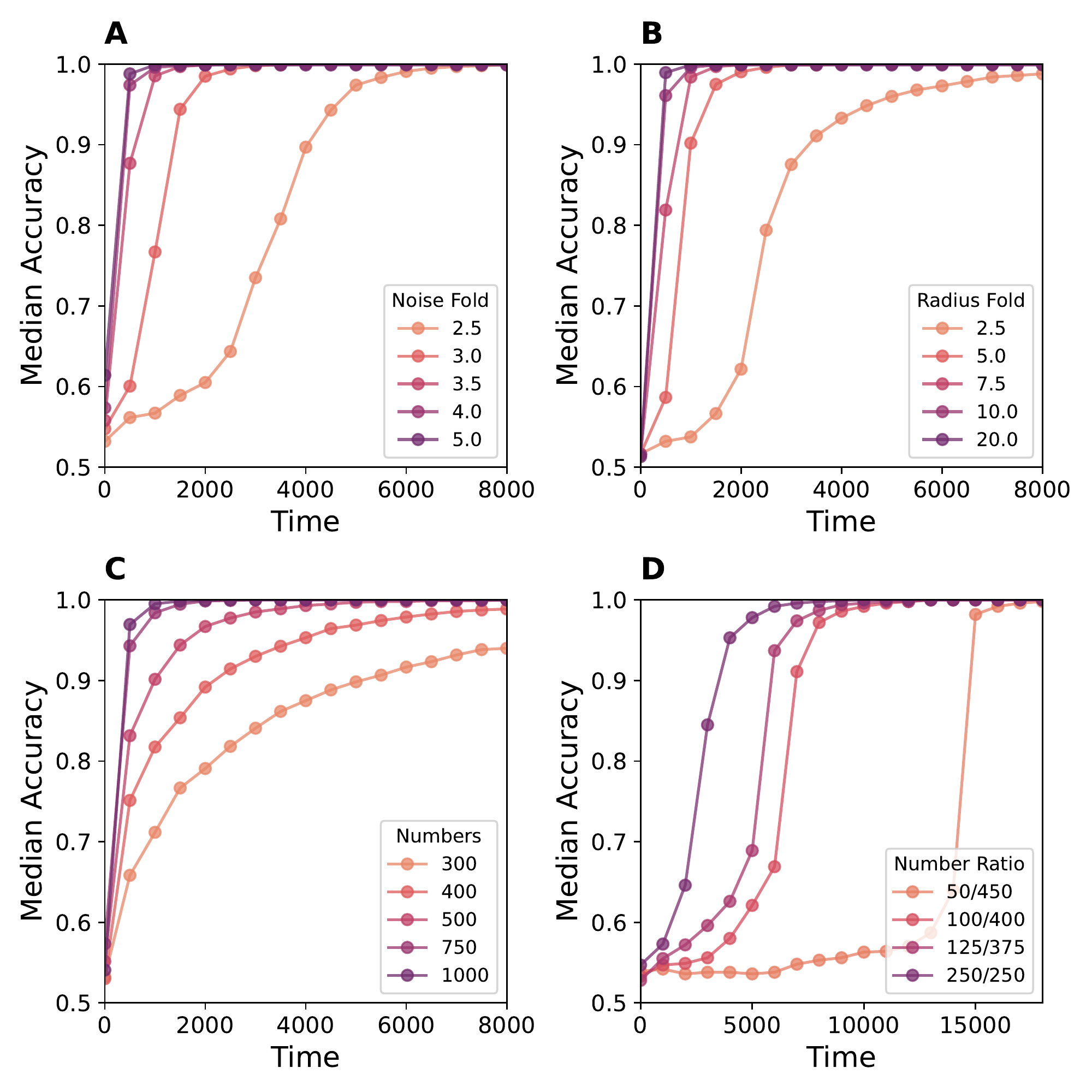}
    \caption{\label{fig:params_plot} \textbf{Parameter influence on timescale of accurate clustering.}
    \boldsf{A}: Median accuracy $N_{\text{sim}}=100$ of clustering for a two sub-species heterogenous Vicsek model with only noise magnitude different. ``Noise fold" refers to the ratio of $\sigma_2/\sigma_1$. \boldsf{B}: Median accuracy clustering two sub-species with only interaction radii different \boldsf{C}: Median accuracy clustering  with the ratio $N_1/N_2=1$  fixed but the total number of particles  $N_1+N_2=N$ is increased. \boldsf{D}: Median accuracy clustering  with the ratio $N_1+N_2=N$ fixed but ratio of two groups is varied. } 
    \end{figure}

The previous result shows that the PC scores in a single collective with two different noise magnitudes cluster over sufficiently long times.  This leaves the natural question of what shapes the timescale for accurate clustering. Due to stochasticity, this timing will differ in each collective. We perform $N_\text{sim}=100$ simulations for each parameter set to evaluate the typical time to cluster accurately for the corresponding scenario. The results of varying the heterogeneity in noise $\sigma$, the interaction radius $R$, and number of particles $N_1, N_2$, and the ratio of $N_1,N_2$ can be seen in \cref{fig:params_plot}.

In \cref{fig:params_plot} panel \boldsf{A}, we see the effect of differing levels of noise between the two subpopulations of particles, ranging from 2.5 to 5.0 ``noise fold", meaning the ratio of $\sigma_2/\sigma_1$. Intuitively, as the populations become more distinct, the ability to distinguish them becomes easier, manifesting as a smaller timescale until all simulations reach 100\% accuracy. In panel \boldsf{B}, a similar effect can be seen for differing only the interaction radii. However, we note that the time for clustering with differing radii takes far longer than clustering noise differences. Next, we investigate the role of particle density by fixing the ratio of $N_1$ to $N_2$ in the noise test of the first panels. We then increase the total number of particles $N=N_1+N_2$ and investigate the time to cluster accurately, finding that the time to cluster decreases with $N$, as seen in panel \boldsf{C}. Lastly, we fix $N$ and vary the ratio of the two subtypes, seen in panel \boldsf{D}. Here, we find that greater asymmetry produces longer accurate clustering time. 
In sum, we find that (i) the more heterogeneous (in parameter values) the subpopulations, (ii) higher density, and (iii) lower asymmetry in numbers all decrease the critical timescale for clustering accurately. 

\begin{figure}[htb]
    \centering
    \includegraphics[width=\linewidth]{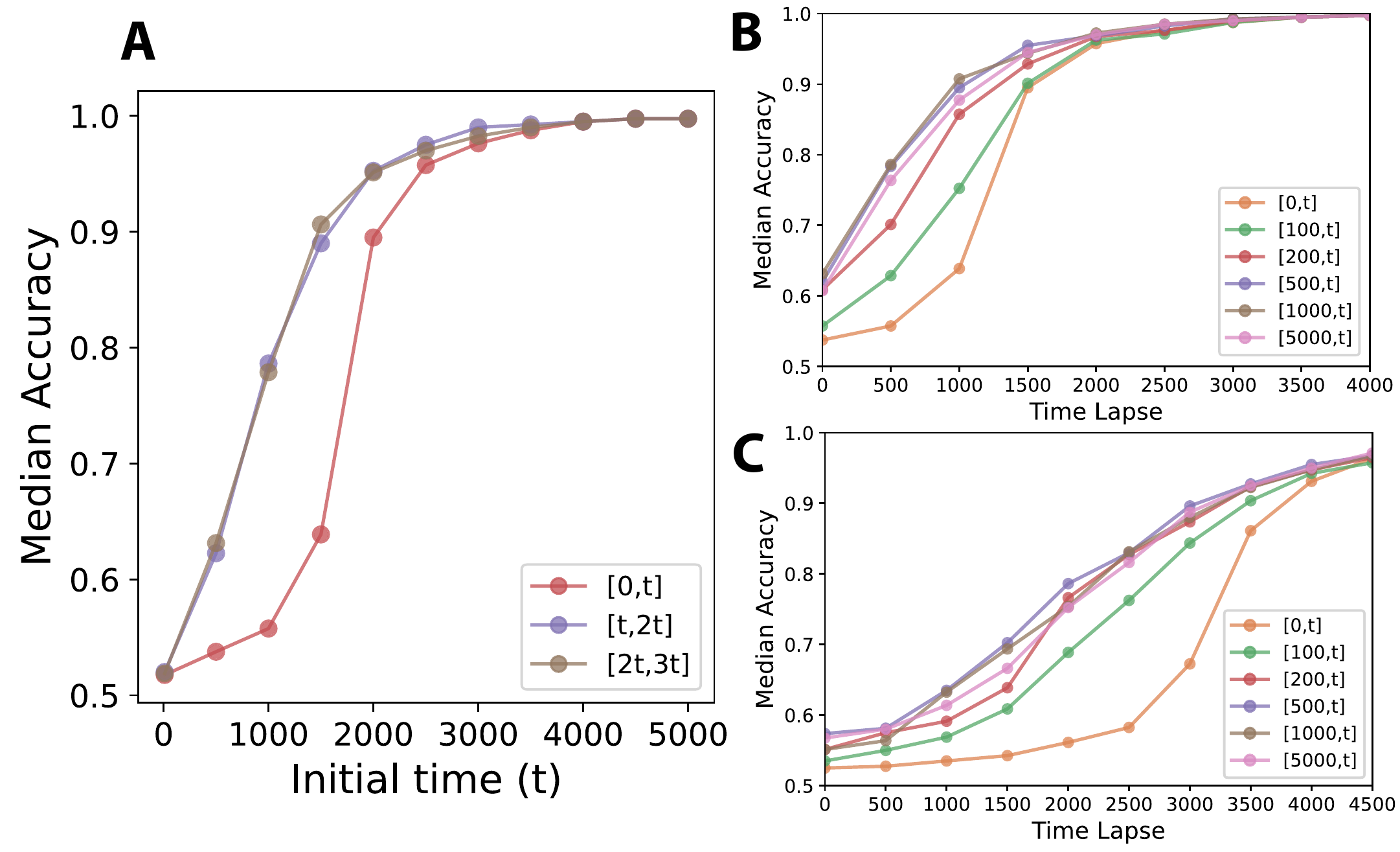}
    \caption{\label{fig:initial}  \edit{\textbf{Clustering timescale dependence transient effects.} All simulations have the same parameters as \cref{fig:vicsek_double} but retain the initial transient timesteps discarded in all other figures.  \boldsf{A}: Median clustering accuracy ( $N_{\mathrm{sim}} = 100$) for sliding windows of time $[0,t]$ (red), $[t,2t]$ (purple), and $[2t,3t]$ showing reduced accuracy for trajectories with transient behavior. \boldsf{B}: Cluster accuracy for various choices of cutoffs for discarding transient effects. Values around $t>500$ converge, supporting the choice of $t=1000$. \boldsf{C}: Same as the previous panel, except with $\sigma_1=0.1,\sigma_2=0.25$, a more challenging clustering ($\sigma_2=0.3$ in \boldsf{B}). Accurate clustering times are longer, but curves for cutoffs $t>200$ appear converged. } }
    \end{figure}

\edit{\subsection{Clustering time is intrinsic and can be disentangled from transient behavior.}}
\label{subsect:transient}

\edit{In the investigation thus far, we have established the intuitive fact that longer trajectories yield higher accuracy in clustering subpopulations. Moreover, the timescale for this accurate clustering is an intrinsic property determined by the parameters of the system. However, it remains unclear whether this emergent timescale is related to transient effects in the system, or corresponds to observations equilibrium. For all simulations unless noted otherwise, we discard the first $t=1000$ time steps in hopes of truly quantifying the equilibrium behavior, but in this section we discuss and investigate this choice. For the heterogeneous two-subpopulation Vicsek model investigated in \cref{fig:params_plot}, we now retain the initial timesteps and denote $t=0$ the initialization with random particle positions and orientations. Then, we investigate sliding windows of time of the trajectories of the same length but at different timepoints. In
Fig. \ref{fig:initial}, panel \boldsf{A}, three curves correspond to clustering accuracy for trajectories limited to $[0,t]$, with no transient effects removed, $[t,2t]$, and $[2t,3t]$. The curve with transient effects is notably distinct from those with initial portions discarded, and has far slower clustering time, with differences occurring on the timescale of $t=1000$. To investigate whether $t=1000$ is an appropriate threshold for cutoff to discard transient timesteps, we vary this threshold and compare the median accuracy for each. In Panel \boldsf{B}, we use the same values as previous figures, including $\sigma_1=0.1$  and $\sigma_2=0.3$, and find that curves with transient times $t>500$ converge onto each other. In panel \boldsf{C}, we further investigate the choice of $t=1000$ cutoff for a harder clustering task $\sigma_1=0.1$ and $\sigma_2=0.25$. While clustering times are broadly longer, all curves for cutoffs $t>200$ again converge onto each other. Altogether, these findings suggest that clustering time is indeed a system-specific emergent quantity even at equilibrium, rather than an artifact of transient behavior.}


\vspace{.1in}

\subsection{More than two subpopulations can be clustered.}

\begin{figure}[htb]
    \centering
    \includegraphics[width=\linewidth]{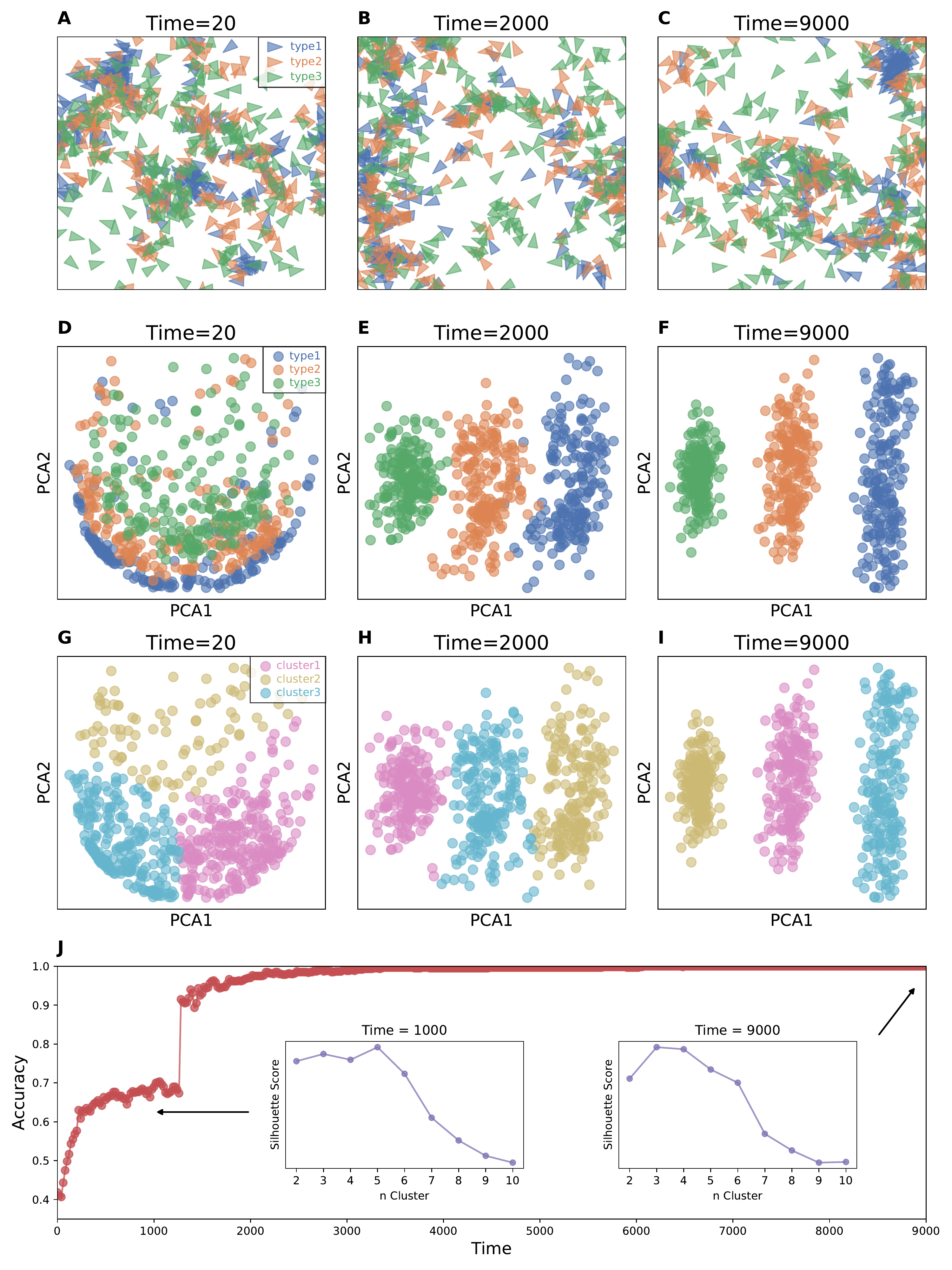}
    \caption{\label{fig:vicsek_triple} \textbf{Three subtype Vicsek model simulation and clustering.}
    \boldsf{ABC}: snapshots of the particle positions in a heterogeneous Vicsek simulation with three types of particles and display no apparent pattern. \boldsf{DEF:} The first two principal component scores for each trajectory, colored by particle type. \boldsf{GHI}: Results of K-means clustering on PC scores. \boldsf{J}: Clustering accuracy approaches 100\% as the trajectories become longer. Inset: silhouette scores at long times correctly identify the number of clusters. The three populations differ only in their noise magnitude $\sigma_1 = 0.1, \sigma_2=0.3, \sigma_3=0.5$ and otherwise $\nu=0.01, R=.05$ with $L=1,\Delta t = 1$, and particle counts$ N_1=200,N_2=200$.}
    \end{figure}
    
The previous examples explore a heterogeneous collective with only two subpopulations. However, the dimensionality reduction and clustering of these latent representations need not be limited to only two populations. We next consider the variation with three subpopulations of Vicsek particles, differing again only by the noise magnitude $\sigma_1=0.1, \sigma_2=0.3,\sigma_3=0.5$. The simulations and clustering procedure can be seen in \cref{fig:vicsek_triple}. Again, the collective itself does not seem to display any apparent pattern in positions (panels \boldsf{ABC}), but the PC scores separate over sufficiently long times (panels \boldsf{DEF})  For long trajectories, the accuracy approaches 100\% (panel \boldsf{J}). In practice, the number of clusters must be specified for K-means or other clustering algorithms but may be unknown. In the inset of panel \boldsf{J}, we plot the silhouette score \cite{sil1987}, a metric for choosing the number of clusters. We see that for intermediate times, an incorrect number of clusters may be inferred (5 clusters is shown as the maximum), but at sufficiently long times, 3 clusters are recovered in the silhouette score as the correct number.

\subsection{Model-free clustering is generalizable to a heterogeneous D'Orsogna model.}

Although the Vicsek has historically served as a testbed for investigations of collective motion, one may wonder whether our results are specific to heterogeneities in this model alone. To explore the generality, we next consider a different, historically important alternative: the D'Orsogna model \cite{dorsogna2006SelfPropelledParticlesSoftCore}.  The D'Orsogna model describes self-propelled particles in 2D, with the position of the $i$th particle $\pmb{x}_i$ evolving  as
\begin{equation}
    \label{eq:dorsogna}
    \frac{\mathrm{d} \pmb{x}_i}{\mathrm{d}t} = \pmb{v}_i, \quad 
    \frac{\mathrm{d} \pmb{v}_i}{\mathrm{d} t} = (\alpha - \beta \|\pmb{v}_i\|^2 )\pmb{v}_i - \nabla U(\pmb{x}_{i}), 
    \end{equation}
where 
    \begin{equation}
    \label{eq:dorsognaMorse}
    U(\pmb{x}_{i}) = \sum_{i\neq j}^N \left[ C_r e^{-\| \pmb{x}_i - \pmb{x}_j \| / \lambda_r} - C_a e^{-\|\pmb{x}_i - \pmb{x}_j\| / \lambda_a} \right]. 
    \end{equation}

In the model, the parameter $\alpha$ describes the self-propulsion magnitude and  $\beta$ is the friction magnitude. The potential  Eq.\eqref{eq:dorsognaMorse} is a Morse-like potential between all pairs of particles. The two length scales are $l_a$ and $l_r$, and represent attraction and repulsion, respectively. Each of those magnitudes is governed by  $C_a$ and $C_r$.  

The D'Orsogna model can display considerably more complex behavior than the Vicsek counterpart. Depending on  the parameter values chosen, possible behaviors range from single mills, double mills, swarms, and escapes \cite{dorsogna2006SelfPropelledParticlesSoftCore,d2019}. Here, we investigate a heterogeneous version of the D'Orsogna model with two subpopulations of particles that each have different parameters \edit{but in a parameter regime where each subpopulation displays the same qualitative behavior. This choice was motivated by the intuition that a setup where subtypes display different qualitative behavior should be easier to cluster and less interesting to investigate.} \edit{We choose $\beta$, the friction, to differ.} The interaction potential sums all neighbors, both in and out of the subtype. One key difference is that the magnitude of the velocity may change in the D'Orsogna model, whereas in Vicsek it is constant. We again use the orientation alone as the data input to the dimensionality reduction, with $\tau_{i,t} = \texttt{atan2}(\pmb{v}^y_{i,t},\pmb{v}^x_{i,t})$ where $\pmb{v}^{x}_{i,t}$ and $\pmb{v}^y_{i,t}$ represent the $x,y$ component of the velocity observed spaced time intervals enumerated by $t$. The ODEs are solved numerically using \texttt{SciPy}'s Dormand-Prince \texttt{dopri5} method and then re-sampled via linear interpolation to be equally spaced observations by $\Delta t=1$.

\begin{figure}[htb]
    \centering
    \includegraphics[width=.99\linewidth]{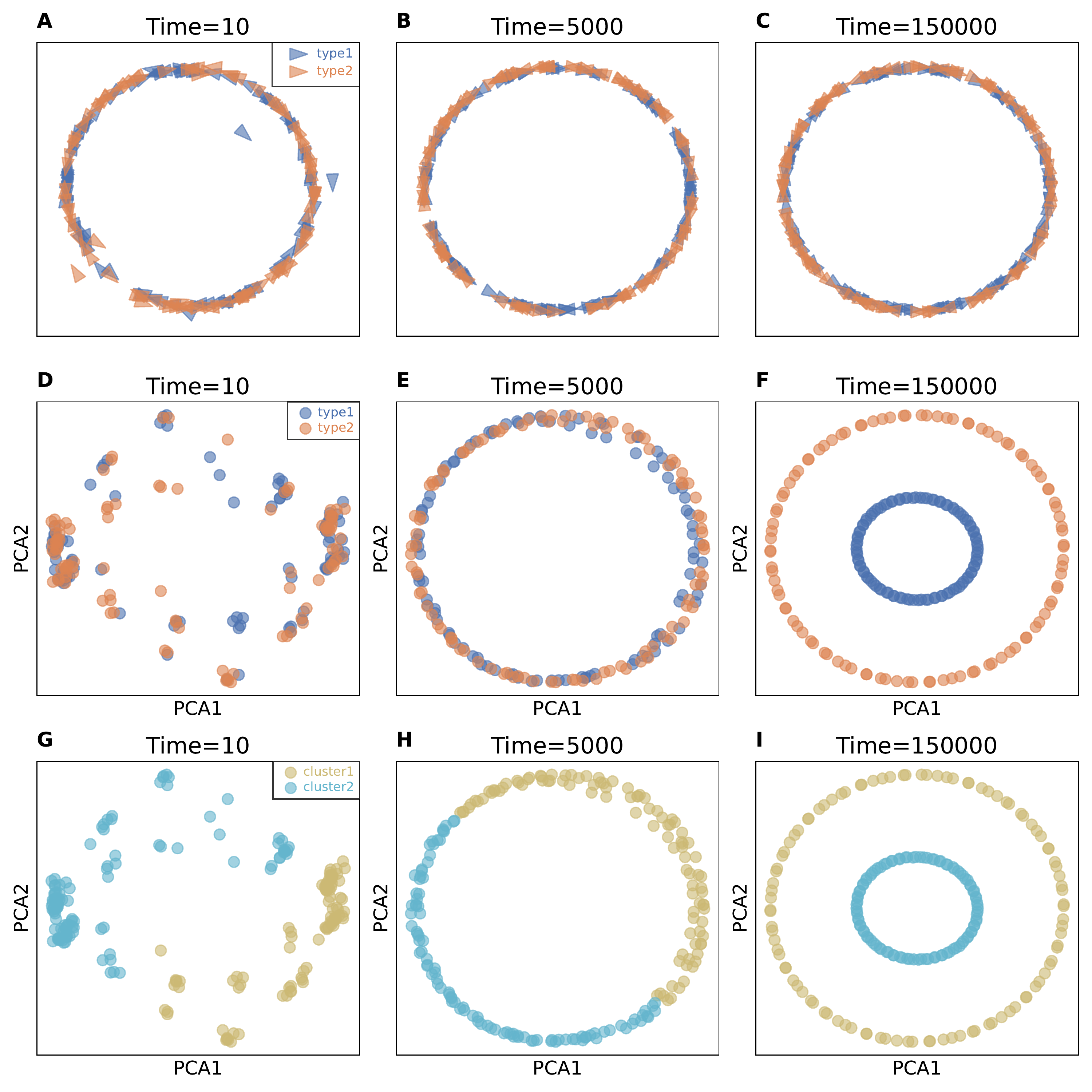}
    \caption{\label{fig:dorsogna} \textbf{Heterogeneous D'Orsogna model simulation and clustering.}
    \boldsf{ABC}: snapshots of the particle positions in a heterogeneous D'Orsogna model simulation with two types of particles and display no apparent pattern. \boldsf{DEF}: The first two principal component scores for each trajectory, colored by particle type. \boldsf{GHI}: Results of spectral clustering on PC scores. Simulation parameter are $N_1=200, N_2=200$ with  shared parameters:  $\alpha=1.50$, $l_a=1.0$, $l_r=0.9$, $C_a=1.0$, $C_r=0.9$, but differing $\beta_1=0.80$ and $\beta_2=0.775$.} 
    \end{figure}

In \cref{eq:dorsogna} we see the results of the heterogeneous D'Orsogna simulation and clustering analysis. For the parameters chosen where attraction is stronger than repulsion, a ring behavior appears with particles moving both clockwise and counterclockwise (\cref{fig:dorsogna} panels \boldsf{ABC}) but otherwise, the identities of each subpopulation do not seem distinguishable by eye. The PC values shown in \boldsf{DEF} do not initially separate the identities, but as longer trajectories are observed, the PC scores from each subtype separate into two circles: those in type 1 with a smaller radius. Due to the shape of the PC scores, K-means expectedly fails to recover the true identities, but standard spectral clustering~\cite{spec2007} shown in \boldsf{GHI} recovers the true identities with flawless accuracy.

\subsection{Limitations on multiple datasets}

\begin{figure}[htb]
    \centering
    \includegraphics[width=.99\linewidth]{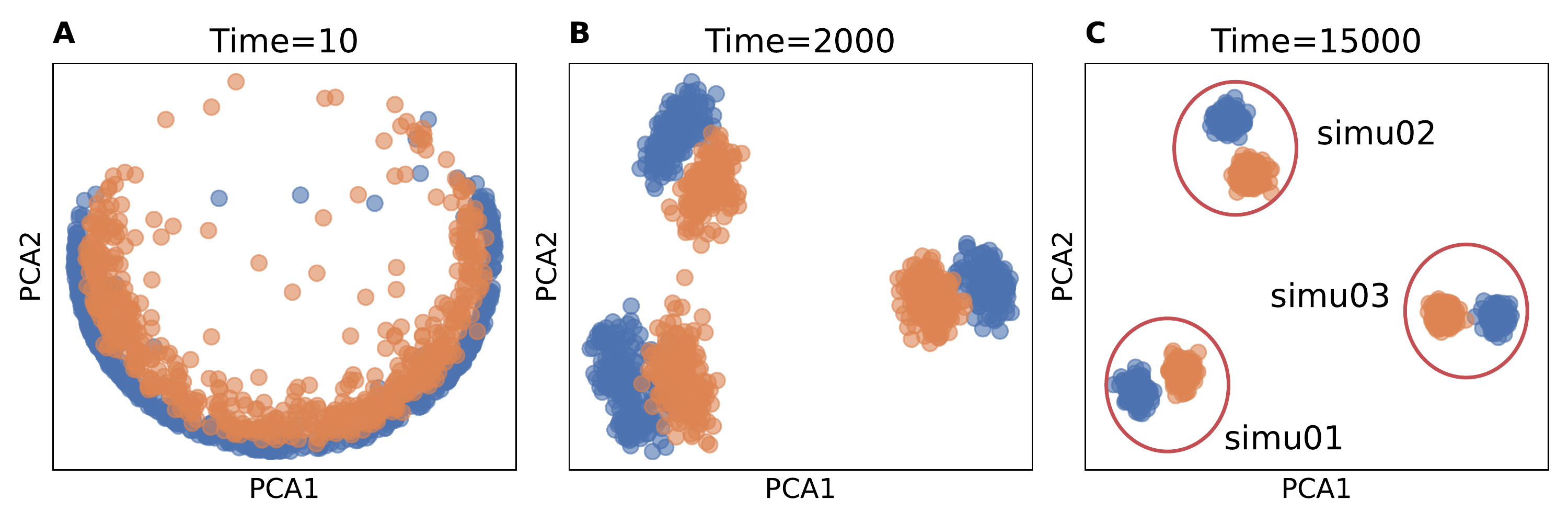}
    \caption{\label{fig:multi_simu} 
    \textbf{Clustering fails to combine multiple experiments.}  
    \boldsf{ABC}: PC scores over increasingly long trajectories with three separate collectives concatenated into a single dataset. 
 The same setup of two-subpopulation Vicsek with different noises, as in \cref{fig:vicsek_double}.}
    \end{figure}

We have thus far investigated the ability to interrogate a single collective at a time and found that we need sufficiently long trajectories for accurate clustering. However, in practice, experimental constraints limit the ability to take long observations. Instead, it may be more practical to obtain replicates of experiments. We therefore investigate the feasibility of combining data from multiple distinct observations of the same heterogeneous collective. Returning to the setup with two subpopulations of Vicsek particles with differing noise magnitude with run $N_1=200, N_2=200$, as in \cref{fig:vicsek_double}, we now run 3 separate simulations. \edit{Each simulation is initialized with different random configurations, and then run to steady-state with these transient values discarded, the same as previous figures.} The three simulations are concatenated into $3 \times 400$ trajectories in one data matrix to cluster. The resulting PC values for the concatenated data can be seen in \cref{fig:multi_simu}. At short times, no apparent pattern is seen. As time progresses (panel \boldsf{B}), the PC scores split into 3 groups. This pattern continues at long times (panel \boldsf{C}), and each of the 3 groups splits into 2 subgroups, resulting in 6 total clusters. However, the 3 predominant groups correspond to the 3 distinct simulations. Therefore, the clustering distinguishes different simulations rather than the same groups between simulations. That is, there does not appear to be a way to tell from the PC scores alone that the 3 observations were from the same heterogeneous collectives. \edit{Intuitively, this is because the temporal structures (correlations) that allow for the statistical separation are limited to a single observation. However, this does not mean the task of identification across multiple experiments is impossible, but rather that it seems to require different techniques that incorporate model structure e.g., the mixture modeling of \cite{Messenger2022}.}

\section{Conclusion}

In summary, we have investigated the ability to perform clustering to recover the true identities of particles in heterogeneous collectives without prior knowledge of the heterogeneities or underlying model. To do so, we first investigated a heterogeneous Vicsek model. To cluster, the orientations are transformed to non-angular data and then dimensionally reduced via PCA. In these latent dimensions, we find that the trajectories naturally separate over sufficiently long timescales. We find that this timescale is decreased by larger differences in noise magnitudes, larger differences in interaction radii, higher particle densities, and equal subpopulation numbers. The method was readily extended to a heterogeneous Vicsek setup with three types of particles, where the number of clusters was also recovered via a silhouette score. Finally, we show that the premise also extends to other models of collectives, by investigating a heterogeneous D'Orsogna model. For this model, we find that spectral clustering was necessary due to the complexity of the PCA scores, but these scores also separate distinctly over long time scales. Ultimately, our results add an important vignette to the growing literature on inferring interactions in collectives, especially those with heterogeneities. 

We emphasize that the approach is not intended as an end-all solution to the identification of heterogeneous collectives, but rather complementary to existing approaches. That is, it can be seen as a step of exploratory data analysis to shape the necessary user input to more sophisticated methods such as \cite{luLearningInteractionKernels,Messenger2022,nabeel2022DisentanglingIntrinsicMotion}. One key limitation of our methodology was the inability to identify whether heterogeneities were the same type across different observations. However, the methodology proposed here could be used to identify the existence of heterogeneities and help steer methods such as \cite{Messenger2022,nabeel2023DatadrivenDiscoveryStochastic}, which we anticipate can readily handle learning interactions and assigning identities across observations. 

There are several avenues of future interest stemming from our work, in both the theory and practice of inferring heterogeneous collectives. It would be interesting to compare the performance of dimensionality reduction approaches to disentangling heterogeneities to those based on information-theoretic quantities like transfer entropy \cite{orange2015transfer,butail2016model,mwaffo2017analysis} or Granger causality \cite{fujii2021learning}. The choice of PCA for dimensionality reduction was for simplicity, but future work could also investigate the use of nonlinear approaches such as autoencoders~\cite{wang2016auto} or LSTM architectures~\cite{yu2019review}. Further, our investigation of heterogeneous collectives was purely numerical. It is therefore of clear interest to explore whether powerful analytical approaches (e.g., Toner-Tu theory \cite{toner1998flocks}) can reveal the intrinsic lower dimensional structure of these heterogeneous collectives. \edit{We emphasize the plausibility of future analytical progress by noting the appearance of clusters from a single principal component, effectively the covariance between the positions of particles.} Such lower dimensional structures have been analytically derived elsewhere for noisy interacting systems \cite{zagli2023DimensionReductionNoisy}, and may reveal further insights about the nature of intrinsic disentanglement of heterogeneities we investigate in this work.

\section{Acknowledgements}
We thank  Mengjian Hua,  Laura Sun, and Mian Wang for working on early preliminaries of this project while undergraduates at New York University.

\section{Code Availability}
Python code for performing the simulations of the heterogeneous collectives and the clustering analysis therefore can be found at \url{https://github.com/tanpei0513/vicsek_trajectory}.

\renewcommand{\bibfont}{\normalfont\scriptsize}
\bibliography{collective}

\begin{thebibliography}{74}%
\makeatletter
\providecommand \@ifxundefined [1]{%
 \@ifx{#1\undefined}
}%
\providecommand \@ifnum [1]{%
 \ifnum #1\expandafter \@firstoftwo
 \else \expandafter \@secondoftwo
 \fi
}%
\providecommand \@ifx [1]{%
 \ifx #1\expandafter \@firstoftwo
 \else \expandafter \@secondoftwo
 \fi
}%
\providecommand \natexlab [1]{#1}%
\providecommand \enquote  [1]{``#1''}%
\providecommand \bibnamefont  [1]{#1}%
\providecommand \bibfnamefont [1]{#1}%
\providecommand \citenamefont [1]{#1}%
\providecommand \href@noop [0]{\@secondoftwo}%
\providecommand \href [0]{\begingroup \@sanitize@url \@href}%
\providecommand \@href[1]{\@@startlink{#1}\@@href}%
\providecommand \@@href[1]{\endgroup#1\@@endlink}%
\providecommand \@sanitize@url [0]{\catcode `\\12\catcode `\$12\catcode
  `\&12\catcode `\#12\catcode `\^12\catcode `\_12\catcode `\%12\relax}%
\providecommand \@@startlink[1]{}%
\providecommand \@@endlink[0]{}%
\providecommand \url  [0]{\begingroup\@sanitize@url \@url }%
\providecommand \@url [1]{\endgroup\@href {#1}{\urlprefix }}%
\providecommand \urlprefix  [0]{URL }%
\providecommand \Eprint [0]{\href }%
\providecommand \doibase [0]{https://doi.org/}%
\providecommand \selectlanguage [0]{\@gobble}%
\providecommand \bibinfo  [0]{\@secondoftwo}%
\providecommand \bibfield  [0]{\@secondoftwo}%
\providecommand \translation [1]{[#1]}%
\providecommand \BibitemOpen [0]{}%
\providecommand \bibitemStop [0]{}%
\providecommand \bibitemNoStop [0]{.\EOS\space}%
\providecommand \EOS [0]{\spacefactor3000\relax}%
\providecommand \BibitemShut  [1]{\csname bibitem#1\endcsname}%
\let\auto@bib@innerbib\@empty
\bibitem [{\citenamefont {Vicsek}\ and\ \citenamefont
  {Zafeiris}(2012)}]{vicsek2012collective}%
  \BibitemOpen
  \bibfield  {author} {\bibinfo {author} {\bibfnamefont {T.}~\bibnamefont
  {Vicsek}}\ and\ \bibinfo {author} {\bibfnamefont {A.}~\bibnamefont
  {Zafeiris}},\ }\bibfield  {title} {\bibinfo {title} {Collective motion},\
  }\href {https://doi.org/10.1016/j.physrep.2012.03.004} {\bibfield  {journal}
  {\bibinfo  {journal} {Physics Reports}\ }\textbf {\bibinfo {volume} {517}},\
  \bibinfo {pages} {71} (\bibinfo {year} {2012})}\BibitemShut {NoStop}%
\bibitem [{\citenamefont {Deutsch}\ \emph {et~al.}(2020)\citenamefont
  {Deutsch}, \citenamefont {Friedl}, \citenamefont {Preziosi},\ and\
  \citenamefont {Theraulaz}}]{deutsch2020MultiscaleAnalysisModelling}%
  \BibitemOpen
  \bibfield  {author} {\bibinfo {author} {\bibfnamefont {A.}~\bibnamefont
  {Deutsch}}, \bibinfo {author} {\bibfnamefont {P.}~\bibnamefont {Friedl}},
  \bibinfo {author} {\bibfnamefont {L.}~\bibnamefont {Preziosi}},\ and\
  \bibinfo {author} {\bibfnamefont {G.}~\bibnamefont {Theraulaz}},\ }\bibfield
  {title} {\bibinfo {title} {Multi-scale analysis and modelling of collective
  migration in biological systems},\ }\href
  {https://doi.org/10.1098/rstb.2019.0377} {\bibfield  {journal} {\bibinfo
  {journal} {Philosophical Transactions of the Royal Society B: Biological
  Sciences}\ }\textbf {\bibinfo {volume} {375}},\ \bibinfo {pages} {20190377}
  (\bibinfo {year} {2020})}\BibitemShut {NoStop}%
\bibitem [{\citenamefont {Hubbard}\ \emph {et~al.}(2004)\citenamefont
  {Hubbard}, \citenamefont {Babak}, \citenamefont {Sigurdsson},\ and\
  \citenamefont {Magn{\'u}sson}}]{fish2004}%
  \BibitemOpen
  \bibfield  {author} {\bibinfo {author} {\bibfnamefont {S.}~\bibnamefont
  {Hubbard}}, \bibinfo {author} {\bibfnamefont {P.}~\bibnamefont {Babak}},
  \bibinfo {author} {\bibfnamefont {S.~T.}\ \bibnamefont {Sigurdsson}},\ and\
  \bibinfo {author} {\bibfnamefont {K.~G.}\ \bibnamefont {Magn{\'u}sson}},\
  }\bibfield  {title} {\bibinfo {title} {A model of the formation of fish
  schools and migrations of fish},\ }\href
  {https://doi.org/10.1016/j.ecolmodel.2003.06.006} {\bibfield  {journal}
  {\bibinfo  {journal} {Ecological Modelling}\ }\textbf {\bibinfo {volume}
  {174}},\ \bibinfo {pages} {359} (\bibinfo {year} {2004})}\BibitemShut
  {NoStop}%
\bibitem [{\citenamefont {Jhawar}\ \emph {et~al.}(2020)\citenamefont {Jhawar},
  \citenamefont {Morris}, \citenamefont {{Amith-Kumar}}, \citenamefont
  {Danny~Raj}, \citenamefont {Rogers}, \citenamefont {Rajendran},\ and\
  \citenamefont {Guttal}}]{jhawar2020NoiseinducedSchoolingFish}%
  \BibitemOpen
  \bibfield  {author} {\bibinfo {author} {\bibfnamefont {J.}~\bibnamefont
  {Jhawar}}, \bibinfo {author} {\bibfnamefont {R.~G.}\ \bibnamefont {Morris}},
  \bibinfo {author} {\bibfnamefont {U.~R.}\ \bibnamefont {{Amith-Kumar}}},
  \bibinfo {author} {\bibfnamefont {M.}~\bibnamefont {Danny~Raj}}, \bibinfo
  {author} {\bibfnamefont {T.}~\bibnamefont {Rogers}}, \bibinfo {author}
  {\bibfnamefont {H.}~\bibnamefont {Rajendran}},\ and\ \bibinfo {author}
  {\bibfnamefont {V.}~\bibnamefont {Guttal}},\ }\bibfield  {title} {\bibinfo
  {title} {Noise-induced schooling of fish},\ }\href
  {https://doi.org/10.1038/s41567-020-0787-y} {\bibfield  {journal} {\bibinfo
  {journal} {Nature Physics}\ }\textbf {\bibinfo {volume} {16}},\ \bibinfo
  {pages} {488} (\bibinfo {year} {2020})}\BibitemShut {NoStop}%
\bibitem [{\citenamefont {Bialek}\ \emph {et~al.}(2012)\citenamefont {Bialek},
  \citenamefont {Cavagna}, \citenamefont {Giardina}, \citenamefont {Mora},
  \citenamefont {Silvestri}, \citenamefont {Viale},\ and\ \citenamefont
  {Walczak}}]{bialek2012StatisticalMechanicsNatural}%
  \BibitemOpen
  \bibfield  {author} {\bibinfo {author} {\bibfnamefont {W.}~\bibnamefont
  {Bialek}}, \bibinfo {author} {\bibfnamefont {A.}~\bibnamefont {Cavagna}},
  \bibinfo {author} {\bibfnamefont {I.}~\bibnamefont {Giardina}}, \bibinfo
  {author} {\bibfnamefont {T.}~\bibnamefont {Mora}}, \bibinfo {author}
  {\bibfnamefont {E.}~\bibnamefont {Silvestri}}, \bibinfo {author}
  {\bibfnamefont {M.}~\bibnamefont {Viale}},\ and\ \bibinfo {author}
  {\bibfnamefont {A.~M.}\ \bibnamefont {Walczak}},\ }\bibfield  {title}
  {\bibinfo {title} {Statistical mechanics for natural flocks of birds},\
  }\href {https://doi.org/10.1073/pnas.1118633109} {\bibfield  {journal}
  {\bibinfo  {journal} {Proceedings of the National Academy of Sciences}\
  }\textbf {\bibinfo {volume} {109}},\ \bibinfo {pages} {4786} (\bibinfo {year}
  {2012})}\BibitemShut {NoStop}%
\bibitem [{\citenamefont {Ling}\ \emph {et~al.}(2019)\citenamefont {Ling},
  \citenamefont {Mclvor}, \citenamefont {van~der Vaart}, \citenamefont
  {Vaughan}, \citenamefont {Thornton},\ and\ \citenamefont
  {Ouellette}}]{bird2019}%
  \BibitemOpen
  \bibfield  {author} {\bibinfo {author} {\bibfnamefont {H.}~\bibnamefont
  {Ling}}, \bibinfo {author} {\bibfnamefont {G.~E.}\ \bibnamefont {Mclvor}},
  \bibinfo {author} {\bibfnamefont {K.}~\bibnamefont {van~der Vaart}}, \bibinfo
  {author} {\bibfnamefont {R.~T.}\ \bibnamefont {Vaughan}}, \bibinfo {author}
  {\bibfnamefont {A.}~\bibnamefont {Thornton}},\ and\ \bibinfo {author}
  {\bibfnamefont {N.~T.}\ \bibnamefont {Ouellette}},\ }\bibfield  {title}
  {\bibinfo {title} {Costs and benefits of social relationships in the
  collective motion of bird flocks},\ }\href
  {https://doi.org/10.1038/s41559-019-0891-5} {\bibfield  {journal} {\bibinfo
  {journal} {Nature ecology \& evolution}\ }\textbf {\bibinfo {volume} {3}},\
  \bibinfo {pages} {943} (\bibinfo {year} {2019})}\BibitemShut {NoStop}%
\bibitem [{\citenamefont {Bernoff}\ \emph {et~al.}(2020)\citenamefont
  {Bernoff}, \citenamefont {{Culshaw-Maurer}}, \citenamefont {Everett},
  \citenamefont {Hohn}, \citenamefont {Strickland},\ and\ \citenamefont
  {Weinburd}}]{bernoff2020AgentbasedContinuousModels}%
  \BibitemOpen
  \bibfield  {author} {\bibinfo {author} {\bibfnamefont {A.~J.}\ \bibnamefont
  {Bernoff}}, \bibinfo {author} {\bibfnamefont {M.}~\bibnamefont
  {{Culshaw-Maurer}}}, \bibinfo {author} {\bibfnamefont {R.~A.}\ \bibnamefont
  {Everett}}, \bibinfo {author} {\bibfnamefont {M.~E.}\ \bibnamefont {Hohn}},
  \bibinfo {author} {\bibfnamefont {W.~C.}\ \bibnamefont {Strickland}},\ and\
  \bibinfo {author} {\bibfnamefont {J.}~\bibnamefont {Weinburd}},\ }\bibfield
  {title} {\bibinfo {title} {Agent-based and continuous models of hopper bands
  for the {{Australian}} plague locust: {{How}} resource consumption mediates
  pulse formation and geometry},\ }\href
  {https://doi.org/10.1371/journal.pcbi.1007820} {\bibfield  {journal}
  {\bibinfo  {journal} {PLOS Computational Biology}\ }\textbf {\bibinfo
  {volume} {16}},\ \bibinfo {pages} {e1007820} (\bibinfo {year}
  {2020})}\BibitemShut {NoStop}%
\bibitem [{\citenamefont {Weinburd}\ \emph {et~al.}(2021)\citenamefont
  {Weinburd}, \citenamefont {Landsberg}, \citenamefont {Kravtsova},
  \citenamefont {Lam}, \citenamefont {Sharma}, \citenamefont {Simpson},
  \citenamefont {Sword},\ and\ \citenamefont
  {Buhl}}]{weinburd2021AnisotropicInteractionMotion}%
  \BibitemOpen
  \bibfield  {author} {\bibinfo {author} {\bibfnamefont {J.}~\bibnamefont
  {Weinburd}}, \bibinfo {author} {\bibfnamefont {J.}~\bibnamefont {Landsberg}},
  \bibinfo {author} {\bibfnamefont {A.}~\bibnamefont {Kravtsova}}, \bibinfo
  {author} {\bibfnamefont {S.}~\bibnamefont {Lam}}, \bibinfo {author}
  {\bibfnamefont {T.}~\bibnamefont {Sharma}}, \bibinfo {author} {\bibfnamefont
  {S.~J.}\ \bibnamefont {Simpson}}, \bibinfo {author} {\bibfnamefont {G.~A.}\
  \bibnamefont {Sword}},\ and\ \bibinfo {author} {\bibfnamefont
  {J.}~\bibnamefont {Buhl}},\ }\href
  {https://doi.org/10.1101/2021.10.29.466390} {\emph {\bibinfo {title}
  {Anisotropic {{Interaction}} and {{Motion States}} of {{Locusts}} in a
  {{Hopper Band}}}}},\ \bibinfo {type} {Preprint}\ (\bibinfo  {institution}
  {{Animal Behavior and Cognition}},\ \bibinfo {year} {2021})\BibitemShut
  {NoStop}%
\bibitem [{\citenamefont {Zhang}\ \emph {et~al.}(2010)\citenamefont {Zhang},
  \citenamefont {Be'er}, \citenamefont {Florin},\ and\ \citenamefont
  {Swinney}}]{bacterial2010}%
  \BibitemOpen
  \bibfield  {author} {\bibinfo {author} {\bibfnamefont {H.-P.}\ \bibnamefont
  {Zhang}}, \bibinfo {author} {\bibfnamefont {A.}~\bibnamefont {Be'er}},
  \bibinfo {author} {\bibfnamefont {E.-L.}\ \bibnamefont {Florin}},\ and\
  \bibinfo {author} {\bibfnamefont {H.~L.}\ \bibnamefont {Swinney}},\
  }\bibfield  {title} {\bibinfo {title} {Collective motion and density
  fluctuations in bacterial colonies},\ }\href
  {https://doi.org/10.1073/pnas.1001651107} {\bibfield  {journal} {\bibinfo
  {journal} {Proceedings of the National Academy of Sciences}\ }\textbf
  {\bibinfo {volume} {107}},\ \bibinfo {pages} {13626} (\bibinfo {year}
  {2010})}\BibitemShut {NoStop}%
\bibitem [{\citenamefont {Peruani}\ \emph {et~al.}(2012)\citenamefont
  {Peruani}, \citenamefont {Starru{\ss}}, \citenamefont {Jakovljevic},
  \citenamefont {{S{\o}gaard-Andersen}}, \citenamefont {Deutsch},\ and\
  \citenamefont {B{\"a}r}}]{bacterial2012}%
  \BibitemOpen
  \bibfield  {author} {\bibinfo {author} {\bibfnamefont {F.}~\bibnamefont
  {Peruani}}, \bibinfo {author} {\bibfnamefont {J.}~\bibnamefont
  {Starru{\ss}}}, \bibinfo {author} {\bibfnamefont {V.}~\bibnamefont
  {Jakovljevic}}, \bibinfo {author} {\bibfnamefont {L.}~\bibnamefont
  {{S{\o}gaard-Andersen}}}, \bibinfo {author} {\bibfnamefont {A.}~\bibnamefont
  {Deutsch}},\ and\ \bibinfo {author} {\bibfnamefont {M.}~\bibnamefont
  {B{\"a}r}},\ }\bibfield  {title} {\bibinfo {title} {Collective motion and
  nonequilibrium cluster formation in colonies of gliding bacteria},\ }\href
  {https://doi.org/10.1103/PhysRevLett.108.098102} {\bibfield  {journal}
  {\bibinfo  {journal} {Physical Review Letters}\ }\textbf {\bibinfo {volume}
  {108}},\ \bibinfo {pages} {098102} (\bibinfo {year} {2012})}\BibitemShut
  {NoStop}%
\bibitem [{\citenamefont {Rio}\ \emph {et~al.}(2018)\citenamefont {Rio},
  \citenamefont {Dachner},\ and\ \citenamefont
  {Warren}}]{rio2018LocalInteractionsUnderlying}%
  \BibitemOpen
  \bibfield  {author} {\bibinfo {author} {\bibfnamefont {K.~W.}\ \bibnamefont
  {Rio}}, \bibinfo {author} {\bibfnamefont {G.~C.}\ \bibnamefont {Dachner}},\
  and\ \bibinfo {author} {\bibfnamefont {W.~H.}\ \bibnamefont {Warren}},\
  }\bibfield  {title} {\bibinfo {title} {Local interactions underlying
  collective motion in human crowds},\ }\href
  {https://doi.org/10.1098/rspb.2018.0611} {\bibfield  {journal} {\bibinfo
  {journal} {Proceedings of the Royal Society B: Biological Sciences}\ }\textbf
  {\bibinfo {volume} {285}},\ \bibinfo {pages} {20180611} (\bibinfo {year}
  {2018})}\BibitemShut {NoStop}%
\bibitem [{\citenamefont {M{\'e}hes}\ and\ \citenamefont
  {Vicsek}(2014)}]{vicsek2014}%
  \BibitemOpen
  \bibfield  {author} {\bibinfo {author} {\bibfnamefont {E.}~\bibnamefont
  {M{\'e}hes}}\ and\ \bibinfo {author} {\bibfnamefont {T.}~\bibnamefont
  {Vicsek}},\ }\bibfield  {title} {\bibinfo {title} {Collective motion of
  cells: From experiments to models},\ }\href
  {https://doi.org/10.1039/c4ib00115j} {\bibfield  {journal} {\bibinfo
  {journal} {Integrative Biology}\ }\textbf {\bibinfo {volume} {6}},\ \bibinfo
  {pages} {831} (\bibinfo {year} {2014})}\BibitemShut {NoStop}%
\bibitem [{\citenamefont {Alert}\ and\ \citenamefont
  {Trepat}(2020)}]{migration2020}%
  \BibitemOpen
  \bibfield  {author} {\bibinfo {author} {\bibfnamefont {R.}~\bibnamefont
  {Alert}}\ and\ \bibinfo {author} {\bibfnamefont {X.}~\bibnamefont {Trepat}},\
  }\bibfield  {title} {\bibinfo {title} {Physical models of collective cell
  migration},\ }\href
  {https://doi.org/10.1146/annurev-conmatphys-031218-013516} {\bibfield
  {journal} {\bibinfo  {journal} {Annual Review of Condensed Matter Physics}\
  }\textbf {\bibinfo {volume} {11}},\ \bibinfo {pages} {77} (\bibinfo {year}
  {2020})}\BibitemShut {NoStop}%
\bibitem [{\citenamefont {Schaller}\ \emph {et~al.}(2010)\citenamefont
  {Schaller}, \citenamefont {Weber}, \citenamefont {Semmrich}, \citenamefont
  {Frey},\ and\ \citenamefont {Bausch}}]{schaller2010polar}%
  \BibitemOpen
  \bibfield  {author} {\bibinfo {author} {\bibfnamefont {V.}~\bibnamefont
  {Schaller}}, \bibinfo {author} {\bibfnamefont {C.}~\bibnamefont {Weber}},
  \bibinfo {author} {\bibfnamefont {C.}~\bibnamefont {Semmrich}}, \bibinfo
  {author} {\bibfnamefont {E.}~\bibnamefont {Frey}},\ and\ \bibinfo {author}
  {\bibfnamefont {A.~R.}\ \bibnamefont {Bausch}},\ }\bibfield  {title}
  {\bibinfo {title} {Polar patterns of driven filaments},\ }\href
  {https://doi.org/10.1038/nature09312} {\bibfield  {journal} {\bibinfo
  {journal} {Nature}\ }\textbf {\bibinfo {volume} {467}},\ \bibinfo {pages}
  {73} (\bibinfo {year} {2010})}\BibitemShut {NoStop}%
\bibitem [{\citenamefont {Miles}\ \emph {et~al.}(2022)\citenamefont {Miles},
  \citenamefont {Zhu},\ and\ \citenamefont
  {Mogilner}}]{miles2022MechanicalTorquePromotes}%
  \BibitemOpen
  \bibfield  {author} {\bibinfo {author} {\bibfnamefont {C.~E.}\ \bibnamefont
  {Miles}}, \bibinfo {author} {\bibfnamefont {J.}~\bibnamefont {Zhu}},\ and\
  \bibinfo {author} {\bibfnamefont {A.}~\bibnamefont {Mogilner}},\ }\bibfield
  {title} {\bibinfo {title} {Mechanical {{Torque Promotes Bipolarity}} of the
  {{Mitotic Spindle Through Multi-centrosomal Clustering}}},\ }\href
  {https://doi.org/10.1007/s11538-021-00985-2} {\bibfield  {journal} {\bibinfo
  {journal} {Bulletin of Mathematical Biology}\ }\textbf {\bibinfo {volume}
  {84}},\ \bibinfo {pages} {29} (\bibinfo {year} {2022})}\BibitemShut {NoStop}%
\bibitem [{\citenamefont {Jolles}\ \emph {et~al.}(2020)\citenamefont {Jolles},
  \citenamefont {King},\ and\ \citenamefont {Killen}}]{jolles2020role}%
  \BibitemOpen
  \bibfield  {author} {\bibinfo {author} {\bibfnamefont {J.~W.}\ \bibnamefont
  {Jolles}}, \bibinfo {author} {\bibfnamefont {A.~J.}\ \bibnamefont {King}},\
  and\ \bibinfo {author} {\bibfnamefont {S.~S.}\ \bibnamefont {Killen}},\
  }\bibfield  {title} {\bibinfo {title} {The role of individual heterogeneity
  in collective animal behaviour},\ }\href
  {https://doi.org/10.1016/j.tree.2019.11.001} {\bibfield  {journal} {\bibinfo
  {journal} {Trends in ecology \& evolution}\ }\textbf {\bibinfo {volume}
  {35}},\ \bibinfo {pages} {278} (\bibinfo {year} {2020})}\BibitemShut
  {NoStop}%
\bibitem [{\citenamefont {Ariel}\ \emph {et~al.}(2022)\citenamefont {Ariel},
  \citenamefont {Ayali}, \citenamefont {Be’er},\ and\ \citenamefont
  {Knebel}}]{ariel2022variability}%
  \BibitemOpen
  \bibfield  {author} {\bibinfo {author} {\bibfnamefont {G.}~\bibnamefont
  {Ariel}}, \bibinfo {author} {\bibfnamefont {A.}~\bibnamefont {Ayali}},
  \bibinfo {author} {\bibfnamefont {A.}~\bibnamefont {Be’er}},\ and\ \bibinfo
  {author} {\bibfnamefont {D.}~\bibnamefont {Knebel}},\ }\bibfield  {title}
  {\bibinfo {title} {Variability and heterogeneity in natural swarms:
  Experiments and modeling},\ }in\ \href
  {https://doi.org/10.1007/978-3-030-93302-9_1} {\emph {\bibinfo {booktitle}
  {Active Particles, Volume 3: Advances in Theory, Models, and Applications}}}\
  (\bibinfo  {publisher} {Springer},\ \bibinfo {year} {2022})\ pp.\ \bibinfo
  {pages} {1--33}\BibitemShut {NoStop}%
\bibitem [{\citenamefont {Peled}\ \emph {et~al.}(2021)\citenamefont {Peled},
  \citenamefont {Ryan}, \citenamefont {Heidenreich}, \citenamefont {B{\"a}r},
  \citenamefont {Ariel},\ and\ \citenamefont
  {Be'er}}]{peled2021HeterogeneousBacterialSwarms}%
  \BibitemOpen
  \bibfield  {author} {\bibinfo {author} {\bibfnamefont {S.}~\bibnamefont
  {Peled}}, \bibinfo {author} {\bibfnamefont {S.~D.}\ \bibnamefont {Ryan}},
  \bibinfo {author} {\bibfnamefont {S.}~\bibnamefont {Heidenreich}}, \bibinfo
  {author} {\bibfnamefont {M.}~\bibnamefont {B{\"a}r}}, \bibinfo {author}
  {\bibfnamefont {G.}~\bibnamefont {Ariel}},\ and\ \bibinfo {author}
  {\bibfnamefont {A.}~\bibnamefont {Be'er}},\ }\bibfield  {title} {\bibinfo
  {title} {Heterogeneous bacterial swarms with mixed lengths},\ }\href
  {https://doi.org/10.1103/PhysRevE.103.032413} {\bibfield  {journal} {\bibinfo
   {journal} {Physical Review E}\ }\textbf {\bibinfo {volume} {103}},\ \bibinfo
  {pages} {032413} (\bibinfo {year} {2021})}\BibitemShut {NoStop}%
\bibitem [{\citenamefont {Ward}\ \emph {et~al.}(2018)\citenamefont {Ward},
  \citenamefont {Schaerf}, \citenamefont {Burns}, \citenamefont {Lizier},
  \citenamefont {Crosato}, \citenamefont {Prokopenko},\ and\ \citenamefont
  {Webster}}]{ward2018cohesion}%
  \BibitemOpen
  \bibfield  {author} {\bibinfo {author} {\bibfnamefont {A.~J.}\ \bibnamefont
  {Ward}}, \bibinfo {author} {\bibfnamefont {T.}~\bibnamefont {Schaerf}},
  \bibinfo {author} {\bibfnamefont {A.}~\bibnamefont {Burns}}, \bibinfo
  {author} {\bibfnamefont {J.}~\bibnamefont {Lizier}}, \bibinfo {author}
  {\bibfnamefont {E.}~\bibnamefont {Crosato}}, \bibinfo {author} {\bibfnamefont
  {M.}~\bibnamefont {Prokopenko}},\ and\ \bibinfo {author} {\bibfnamefont
  {M.~M.}\ \bibnamefont {Webster}},\ }\bibfield  {title} {\bibinfo {title}
  {Cohesion, order and information flow in the collective motion of
  mixed-species shoals},\ }\href {https://doi.org/doi.org/10.1098/rsos.181132}
  {\bibfield  {journal} {\bibinfo  {journal} {Royal Society Open Science}\
  }\textbf {\bibinfo {volume} {5}},\ \bibinfo {pages} {181132} (\bibinfo {year}
  {2018})}\BibitemShut {NoStop}%
\bibitem [{\citenamefont {{Herbert-Read}}\ \emph {et~al.}(2013)\citenamefont
  {{Herbert-Read}}, \citenamefont {Krause}, \citenamefont {Morrell},
  \citenamefont {Schaerf}, \citenamefont {Krause},\ and\ \citenamefont
  {Ward}}]{herbert-read2013RoleIndividualityCollective}%
  \BibitemOpen
  \bibfield  {author} {\bibinfo {author} {\bibfnamefont {J.~E.}\ \bibnamefont
  {{Herbert-Read}}}, \bibinfo {author} {\bibfnamefont {S.}~\bibnamefont
  {Krause}}, \bibinfo {author} {\bibfnamefont {L.~J.}\ \bibnamefont {Morrell}},
  \bibinfo {author} {\bibfnamefont {T.~M.}\ \bibnamefont {Schaerf}}, \bibinfo
  {author} {\bibfnamefont {J.}~\bibnamefont {Krause}},\ and\ \bibinfo {author}
  {\bibfnamefont {A.~J.~W.}\ \bibnamefont {Ward}},\ }\bibfield  {title}
  {\bibinfo {title} {The role of individuality in collective group movement},\
  }\href {https://doi.org/10.1098/rspb.2012.2564} {\bibfield  {journal}
  {\bibinfo  {journal} {Proceedings of the Royal Society B: Biological
  Sciences}\ }\textbf {\bibinfo {volume} {280}},\ \bibinfo {pages} {20122564}
  (\bibinfo {year} {2013})}\BibitemShut {NoStop}%
\bibitem [{\citenamefont {Collignon}\ \emph {et~al.}(2019)\citenamefont
  {Collignon}, \citenamefont {S{\'e}guret}, \citenamefont {Chemtob},
  \citenamefont {Cazenille},\ and\ \citenamefont
  {Halloy}}]{collignon2019collective}%
  \BibitemOpen
  \bibfield  {author} {\bibinfo {author} {\bibfnamefont {B.}~\bibnamefont
  {Collignon}}, \bibinfo {author} {\bibfnamefont {A.}~\bibnamefont
  {S{\'e}guret}}, \bibinfo {author} {\bibfnamefont {Y.}~\bibnamefont
  {Chemtob}}, \bibinfo {author} {\bibfnamefont {L.}~\bibnamefont {Cazenille}},\
  and\ \bibinfo {author} {\bibfnamefont {J.}~\bibnamefont {Halloy}},\
  }\bibfield  {title} {\bibinfo {title} {Collective departures and leadership
  in zebrafish},\ }\href {https://doi.org/10.1371/journal.pone.0216798}
  {\bibfield  {journal} {\bibinfo  {journal} {PloS ONE}\ }\textbf {\bibinfo
  {volume} {14}},\ \bibinfo {pages} {e0216798} (\bibinfo {year}
  {2019})}\BibitemShut {NoStop}%
\bibitem [{\citenamefont {Mizumoto}\ \emph {et~al.}(2021)\citenamefont
  {Mizumoto}, \citenamefont {Lee}, \citenamefont {Valentini}, \citenamefont
  {Chouvenc},\ and\ \citenamefont
  {Pratt}}]{mizumoto2021CoordinationMovementComplementary}%
  \BibitemOpen
  \bibfield  {author} {\bibinfo {author} {\bibfnamefont {N.}~\bibnamefont
  {Mizumoto}}, \bibinfo {author} {\bibfnamefont {S.-B.}\ \bibnamefont {Lee}},
  \bibinfo {author} {\bibfnamefont {G.}~\bibnamefont {Valentini}}, \bibinfo
  {author} {\bibfnamefont {T.}~\bibnamefont {Chouvenc}},\ and\ \bibinfo
  {author} {\bibfnamefont {S.~C.}\ \bibnamefont {Pratt}},\ }\bibfield  {title}
  {\bibinfo {title} {Coordination of movement via complementary interactions of
  leaders and followers in termite mating pairs},\ }\href
  {https://doi.org/10.1098/rspb.2021.0998} {\bibfield  {journal} {\bibinfo
  {journal} {Proceedings of the Royal Society B: Biological Sciences}\ }\textbf
  {\bibinfo {volume} {288}},\ \bibinfo {pages} {20210998} (\bibinfo {year}
  {2021})}\BibitemShut {NoStop}%
\bibitem [{\citenamefont {{G{\'o}mez-Nava}}\ \emph {et~al.}(2022)\citenamefont
  {{G{\'o}mez-Nava}}, \citenamefont {Bon},\ and\ \citenamefont
  {Peruani}}]{gomez-nava2022IntermittentCollectiveMotion}%
  \BibitemOpen
  \bibfield  {author} {\bibinfo {author} {\bibfnamefont {L.}~\bibnamefont
  {{G{\'o}mez-Nava}}}, \bibinfo {author} {\bibfnamefont {R.}~\bibnamefont
  {Bon}},\ and\ \bibinfo {author} {\bibfnamefont {F.}~\bibnamefont {Peruani}},\
  }\bibfield  {title} {\bibinfo {title} {Intermittent collective motion in
  sheep results from alternating the role of leader and follower},\ }\href
  {https://doi.org/10.1038/s41567-022-01769-8} {\bibfield  {journal} {\bibinfo
  {journal} {Nature Physics}\ }\textbf {\bibinfo {volume} {18}},\ \bibinfo
  {pages} {1494} (\bibinfo {year} {2022})}\BibitemShut {NoStop}%
\bibitem [{\citenamefont {Schumacher}\ \emph {et~al.}(2017)\citenamefont
  {Schumacher}, \citenamefont {Maini},\ and\ \citenamefont
  {Baker}}]{schumacher2017SemblanceHeterogeneityCollective}%
  \BibitemOpen
  \bibfield  {author} {\bibinfo {author} {\bibfnamefont {L.~J.}\ \bibnamefont
  {Schumacher}}, \bibinfo {author} {\bibfnamefont {P.~K.}\ \bibnamefont
  {Maini}},\ and\ \bibinfo {author} {\bibfnamefont {R.~E.}\ \bibnamefont
  {Baker}},\ }\bibfield  {title} {\bibinfo {title} {Semblance of
  {{Heterogeneity}} in {{Collective Cell Migration}}},\ }\href
  {https://doi.org/10.1016/j.cels.2017.06.006} {\bibfield  {journal} {\bibinfo
  {journal} {Cell Systems}\ }\textbf {\bibinfo {volume} {5}},\ \bibinfo {pages}
  {119} (\bibinfo {year} {2017})}\BibitemShut {NoStop}%
\bibitem [{\citenamefont {Fu}\ \emph {et~al.}(2018)\citenamefont {Fu},
  \citenamefont {Kato}, \citenamefont {Long}, \citenamefont {Mattingly},
  \citenamefont {He}, \citenamefont {Vural}, \citenamefont {Zucker},\ and\
  \citenamefont {Emonet}}]{fu2018SpatialSelforganizationResolves}%
  \BibitemOpen
  \bibfield  {author} {\bibinfo {author} {\bibfnamefont {X.}~\bibnamefont
  {Fu}}, \bibinfo {author} {\bibfnamefont {S.}~\bibnamefont {Kato}}, \bibinfo
  {author} {\bibfnamefont {J.}~\bibnamefont {Long}}, \bibinfo {author}
  {\bibfnamefont {H.~H.}\ \bibnamefont {Mattingly}}, \bibinfo {author}
  {\bibfnamefont {C.}~\bibnamefont {He}}, \bibinfo {author} {\bibfnamefont
  {D.~C.}\ \bibnamefont {Vural}}, \bibinfo {author} {\bibfnamefont {S.~W.}\
  \bibnamefont {Zucker}},\ and\ \bibinfo {author} {\bibfnamefont
  {T.}~\bibnamefont {Emonet}},\ }\bibfield  {title} {\bibinfo {title} {Spatial
  self-organization resolves conflicts between individuality and collective
  migration},\ }\href {https://doi.org/10.1038/s41467-018-04539-4} {\bibfield
  {journal} {\bibinfo  {journal} {Nature Communications}\ }\textbf {\bibinfo
  {volume} {9}},\ \bibinfo {pages} {2177} (\bibinfo {year} {2018})}\BibitemShut
  {NoStop}%
\bibitem [{\citenamefont {Kwon}\ \emph {et~al.}(2019)\citenamefont {Kwon},
  \citenamefont {Kwon}, \citenamefont {Cha},\ and\ \citenamefont
  {Sung}}]{kwon2019stochastic}%
  \BibitemOpen
  \bibfield  {author} {\bibinfo {author} {\bibfnamefont {T.}~\bibnamefont
  {Kwon}}, \bibinfo {author} {\bibfnamefont {O.-S.}\ \bibnamefont {Kwon}},
  \bibinfo {author} {\bibfnamefont {H.-J.}\ \bibnamefont {Cha}},\ and\ \bibinfo
  {author} {\bibfnamefont {B.~J.}\ \bibnamefont {Sung}},\ }\bibfield  {title}
  {\bibinfo {title} {Stochastic and heterogeneous cancer cell migration:
  experiment and theory},\ }\href {https://doi.org/10.1038/s41598-019-52480-3}
  {\bibfield  {journal} {\bibinfo  {journal} {Scientific reports}\ }\textbf
  {\bibinfo {volume} {9}},\ \bibinfo {pages} {1} (\bibinfo {year}
  {2019})}\BibitemShut {NoStop}%
\bibitem [{\citenamefont {Qin}\ \emph {et~al.}(2021)\citenamefont {Qin},
  \citenamefont {Yang}, \citenamefont {Yi}, \citenamefont {Cao},\ and\
  \citenamefont {Xiao}}]{qin2021roles}%
  \BibitemOpen
  \bibfield  {author} {\bibinfo {author} {\bibfnamefont {L.}~\bibnamefont
  {Qin}}, \bibinfo {author} {\bibfnamefont {D.}~\bibnamefont {Yang}}, \bibinfo
  {author} {\bibfnamefont {W.}~\bibnamefont {Yi}}, \bibinfo {author}
  {\bibfnamefont {H.}~\bibnamefont {Cao}},\ and\ \bibinfo {author}
  {\bibfnamefont {G.}~\bibnamefont {Xiao}},\ }\bibfield  {title} {\bibinfo
  {title} {Roles of leader and follower cells in collective cell migration},\
  }\href {https://doi.org/10.1091/mbc.E20-10-0681} {\bibfield  {journal}
  {\bibinfo  {journal} {Molecular Biology of the Cell}\ }\textbf {\bibinfo
  {volume} {32}},\ \bibinfo {pages} {1267} (\bibinfo {year}
  {2021})}\BibitemShut {NoStop}%
\bibitem [{\citenamefont {Zhang}\ \emph {et~al.}(2019)\citenamefont {Zhang},
  \citenamefont {Zhu}, \citenamefont {Hostikka},\ and\ \citenamefont
  {Qiu}}]{zhang2019pedestrian}%
  \BibitemOpen
  \bibfield  {author} {\bibinfo {author} {\bibfnamefont {D.}~\bibnamefont
  {Zhang}}, \bibinfo {author} {\bibfnamefont {H.}~\bibnamefont {Zhu}}, \bibinfo
  {author} {\bibfnamefont {S.}~\bibnamefont {Hostikka}},\ and\ \bibinfo
  {author} {\bibfnamefont {S.}~\bibnamefont {Qiu}},\ }\bibfield  {title}
  {\bibinfo {title} {Pedestrian dynamics in a heterogeneous bidirectional flow:
  overtaking behaviour and lane formation},\ }\href
  {https://doi.org/10.1016/j.physa.2019.03.032} {\bibfield  {journal} {\bibinfo
   {journal} {Physica A: Statistical Mechanics and its Applications}\ }\textbf
  {\bibinfo {volume} {525}},\ \bibinfo {pages} {72} (\bibinfo {year}
  {2019})}\BibitemShut {NoStop}%
\bibitem [{\citenamefont {Ariel}\ \emph {et~al.}(2015)\citenamefont {Ariel},
  \citenamefont {Rimer},\ and\ \citenamefont {{Ben-Jacob}}}]{phase2015}%
  \BibitemOpen
  \bibfield  {author} {\bibinfo {author} {\bibfnamefont {G.}~\bibnamefont
  {Ariel}}, \bibinfo {author} {\bibfnamefont {O.}~\bibnamefont {Rimer}},\ and\
  \bibinfo {author} {\bibfnamefont {E.}~\bibnamefont {{Ben-Jacob}}},\
  }\bibfield  {title} {\bibinfo {title} {Order\textendash disorder phase
  transition in heterogeneous populations of self-propelled particles},\ }\href
  {https://doi.org/10.1007/s10955-014-1095-7} {\bibfield  {journal} {\bibinfo
  {journal} {Journal of Statistical Physics}\ }\textbf {\bibinfo {volume}
  {158}},\ \bibinfo {pages} {579} (\bibinfo {year} {2015})}\BibitemShut
  {NoStop}%
\bibitem [{\citenamefont {Copenhagen}\ \emph {et~al.}(2016)\citenamefont
  {Copenhagen}, \citenamefont {Quint},\ and\ \citenamefont
  {Gopinathan}}]{copenhagen2016SelforganizedSortingLimits}%
  \BibitemOpen
  \bibfield  {author} {\bibinfo {author} {\bibfnamefont {K.}~\bibnamefont
  {Copenhagen}}, \bibinfo {author} {\bibfnamefont {D.~A.}\ \bibnamefont
  {Quint}},\ and\ \bibinfo {author} {\bibfnamefont {A.}~\bibnamefont
  {Gopinathan}},\ }\bibfield  {title} {\bibinfo {title} {Self-organized sorting
  limits behavioral variability in swarms},\ }\href
  {https://doi.org/10.1038/srep31808} {\bibfield  {journal} {\bibinfo
  {journal} {Scientific Reports}\ }\textbf {\bibinfo {volume} {6}},\ \bibinfo
  {pages} {31808} (\bibinfo {year} {2016})}\BibitemShut {NoStop}%
\bibitem [{\citenamefont {del Mar~Delgado}\ \emph {et~al.}(2018)\citenamefont
  {del Mar~Delgado}, \citenamefont {Miranda}, \citenamefont {Alvarez},
  \citenamefont {Gurarie}, \citenamefont {Fagan}, \citenamefont {Penteriani},
  \citenamefont {di~Virgilio},\ and\ \citenamefont
  {Morales}}]{del2018importance}%
  \BibitemOpen
  \bibfield  {author} {\bibinfo {author} {\bibfnamefont {M.}~\bibnamefont {del
  Mar~Delgado}}, \bibinfo {author} {\bibfnamefont {M.}~\bibnamefont {Miranda}},
  \bibinfo {author} {\bibfnamefont {S.~J.}\ \bibnamefont {Alvarez}}, \bibinfo
  {author} {\bibfnamefont {E.}~\bibnamefont {Gurarie}}, \bibinfo {author}
  {\bibfnamefont {W.~F.}\ \bibnamefont {Fagan}}, \bibinfo {author}
  {\bibfnamefont {V.}~\bibnamefont {Penteriani}}, \bibinfo {author}
  {\bibfnamefont {A.}~\bibnamefont {di~Virgilio}},\ and\ \bibinfo {author}
  {\bibfnamefont {J.~M.}\ \bibnamefont {Morales}},\ }\bibfield  {title}
  {\bibinfo {title} {The importance of individual variation in the dynamics of
  animal collective movements},\ }\href
  {https://doi.org/10.1098/rstb.2017.0008} {\bibfield  {journal} {\bibinfo
  {journal} {Philosophical Transactions of the Royal Society B: Biological
  Sciences}\ }\textbf {\bibinfo {volume} {373}},\ \bibinfo {pages} {20170008}
  (\bibinfo {year} {2018})}\BibitemShut {NoStop}%
\bibitem [{\citenamefont {Hoell}\ \emph {et~al.}(2019)\citenamefont {Hoell},
  \citenamefont {L{\"o}wen},\ and\ \citenamefont
  {Menzel}}]{hoell2019MultispeciesDynamicalDensity}%
  \BibitemOpen
  \bibfield  {author} {\bibinfo {author} {\bibfnamefont {C.}~\bibnamefont
  {Hoell}}, \bibinfo {author} {\bibfnamefont {H.}~\bibnamefont {L{\"o}wen}},\
  and\ \bibinfo {author} {\bibfnamefont {A.~M.}\ \bibnamefont {Menzel}},\
  }\bibfield  {title} {\bibinfo {title} {Multi-species dynamical density
  functional theory for microswimmers: {{Derivation}}, orientational ordering,
  trapping potentials, and shear cells},\ }\href
  {https://doi.org/10.1063/1.5099554} {\bibfield  {journal} {\bibinfo
  {journal} {The Journal of Chemical Physics}\ }\textbf {\bibinfo {volume}
  {151}},\ \bibinfo {pages} {064902} (\bibinfo {year} {2019})}\BibitemShut
  {NoStop}%
\bibitem [{\citenamefont {Netzer}\ \emph {et~al.}(2019)\citenamefont {Netzer},
  \citenamefont {Yarom},\ and\ \citenamefont
  {Ariel}}]{netzer2019HeterogeneousPopulationsNetwork}%
  \BibitemOpen
  \bibfield  {author} {\bibinfo {author} {\bibfnamefont {G.}~\bibnamefont
  {Netzer}}, \bibinfo {author} {\bibfnamefont {Y.}~\bibnamefont {Yarom}},\ and\
  \bibinfo {author} {\bibfnamefont {G.}~\bibnamefont {Ariel}},\ }\bibfield
  {title} {\bibinfo {title} {Heterogeneous populations in a network model of
  collective motion},\ }\href {https://doi.org/10.1016/j.physa.2019.121550}
  {\bibfield  {journal} {\bibinfo  {journal} {Physica A: Statistical Mechanics
  and its Applications}\ }\textbf {\bibinfo {volume} {530}},\ \bibinfo {pages}
  {121550} (\bibinfo {year} {2019})}\BibitemShut {NoStop}%
\bibitem [{\citenamefont {Khelfa}\ \emph {et~al.}(2022)\citenamefont {Khelfa},
  \citenamefont {Korbmacher}, \citenamefont {Schadschneider},\ and\
  \citenamefont {Tordeux}}]{heter2022}%
  \BibitemOpen
  \bibfield  {author} {\bibinfo {author} {\bibfnamefont {B.}~\bibnamefont
  {Khelfa}}, \bibinfo {author} {\bibfnamefont {R.}~\bibnamefont {Korbmacher}},
  \bibinfo {author} {\bibfnamefont {A.}~\bibnamefont {Schadschneider}},\ and\
  \bibinfo {author} {\bibfnamefont {A.}~\bibnamefont {Tordeux}},\ }\bibfield
  {title} {\bibinfo {title} {Heterogeneity-induced lane and band formation in
  self-driven particle systems},\ }\href
  {https://doi.org/10.1038/s41598-022-08649-4} {\bibfield  {journal} {\bibinfo
  {journal} {Scientific reports}\ }\textbf {\bibinfo {volume} {12}},\ \bibinfo
  {pages} {4768} (\bibinfo {year} {2022})}\BibitemShut {NoStop}%
\bibitem [{\citenamefont {Lukeman}\ \emph {et~al.}(2010)\citenamefont
  {Lukeman}, \citenamefont {Li},\ and\ \citenamefont
  {{Edelstein-Keshet}}}]{lukeman2010InferringIndividualRules}%
  \BibitemOpen
  \bibfield  {author} {\bibinfo {author} {\bibfnamefont {R.}~\bibnamefont
  {Lukeman}}, \bibinfo {author} {\bibfnamefont {Y.-X.}\ \bibnamefont {Li}},\
  and\ \bibinfo {author} {\bibfnamefont {L.}~\bibnamefont
  {{Edelstein-Keshet}}},\ }\bibfield  {title} {\bibinfo {title} {Inferring
  individual rules from collective behavior},\ }\href
  {https://doi.org/10.1073/pnas.1001763107} {\bibfield  {journal} {\bibinfo
  {journal} {Proceedings of the National Academy of Sciences}\ }\textbf
  {\bibinfo {volume} {107}},\ \bibinfo {pages} {12576} (\bibinfo {year}
  {2010})}\BibitemShut {NoStop}%
\bibitem [{\citenamefont {Mann}(2011)}]{mann2011BayesianInferenceIdentifying}%
  \BibitemOpen
  \bibfield  {author} {\bibinfo {author} {\bibfnamefont {R.~P.}\ \bibnamefont
  {Mann}},\ }\bibfield  {title} {\bibinfo {title} {Bayesian {{Inference}} for
  {{Identifying Interaction Rules}} in {{Moving Animal Groups}}},\ }\href
  {https://doi.org/10.1371/journal.pone.0022827} {\bibfield  {journal}
  {\bibinfo  {journal} {PLoS ONE}\ }\textbf {\bibinfo {volume} {6}},\ \bibinfo
  {pages} {e22827} (\bibinfo {year} {2011})}\BibitemShut {NoStop}%
\bibitem [{\citenamefont {{Herbert-Read}}\ \emph {et~al.}(2011)\citenamefont
  {{Herbert-Read}}, \citenamefont {Perna}, \citenamefont {Mann}, \citenamefont
  {Schaerf}, \citenamefont {Sumpter},\ and\ \citenamefont
  {Ward}}]{herbert-read2011InferringRulesInteraction}%
  \BibitemOpen
  \bibfield  {author} {\bibinfo {author} {\bibfnamefont {J.~E.}\ \bibnamefont
  {{Herbert-Read}}}, \bibinfo {author} {\bibfnamefont {A.}~\bibnamefont
  {Perna}}, \bibinfo {author} {\bibfnamefont {R.~P.}\ \bibnamefont {Mann}},
  \bibinfo {author} {\bibfnamefont {T.~M.}\ \bibnamefont {Schaerf}}, \bibinfo
  {author} {\bibfnamefont {D.~J.~T.}\ \bibnamefont {Sumpter}},\ and\ \bibinfo
  {author} {\bibfnamefont {A.~J.~W.}\ \bibnamefont {Ward}},\ }\bibfield
  {title} {\bibinfo {title} {Inferring the rules of interaction of shoaling
  fish},\ }\href {https://doi.org/10.1073/pnas.1109355108} {\bibfield
  {journal} {\bibinfo  {journal} {Proceedings of the National Academy of
  Sciences}\ }\textbf {\bibinfo {volume} {108}},\ \bibinfo {pages} {18726}
  (\bibinfo {year} {2011})}\BibitemShut {NoStop}%
\bibitem [{\citenamefont {Katz}\ \emph {et~al.}(2011)\citenamefont {Katz},
  \citenamefont {Tunstr{\o}m}, \citenamefont {Ioannou}, \citenamefont {Huepe},\
  and\ \citenamefont {Couzin}}]{katz2011InferringStructureDynamics}%
  \BibitemOpen
  \bibfield  {author} {\bibinfo {author} {\bibfnamefont {Y.}~\bibnamefont
  {Katz}}, \bibinfo {author} {\bibfnamefont {K.}~\bibnamefont {Tunstr{\o}m}},
  \bibinfo {author} {\bibfnamefont {C.~C.}\ \bibnamefont {Ioannou}}, \bibinfo
  {author} {\bibfnamefont {C.}~\bibnamefont {Huepe}},\ and\ \bibinfo {author}
  {\bibfnamefont {I.~D.}\ \bibnamefont {Couzin}},\ }\bibfield  {title}
  {\bibinfo {title} {Inferring the structure and dynamics of interactions in
  schooling fish},\ }\href {https://doi.org/10.1073/pnas.1107583108} {\bibfield
   {journal} {\bibinfo  {journal} {Proceedings of the National Academy of
  Sciences}\ }\textbf {\bibinfo {volume} {108}},\ \bibinfo {pages} {18720}
  (\bibinfo {year} {2011})}\BibitemShut {NoStop}%
\bibitem [{\citenamefont {Gautrais}\ \emph {et~al.}(2012)\citenamefont
  {Gautrais}, \citenamefont {Ginelli}, \citenamefont {Fournier}, \citenamefont
  {Blanco}, \citenamefont {Soria}, \citenamefont {Chat{\'e}},\ and\
  \citenamefont {Theraulaz}}]{gautrais2012DecipheringInteractionsMoving}%
  \BibitemOpen
  \bibfield  {author} {\bibinfo {author} {\bibfnamefont {J.}~\bibnamefont
  {Gautrais}}, \bibinfo {author} {\bibfnamefont {F.}~\bibnamefont {Ginelli}},
  \bibinfo {author} {\bibfnamefont {R.}~\bibnamefont {Fournier}}, \bibinfo
  {author} {\bibfnamefont {S.}~\bibnamefont {Blanco}}, \bibinfo {author}
  {\bibfnamefont {M.}~\bibnamefont {Soria}}, \bibinfo {author} {\bibfnamefont
  {H.}~\bibnamefont {Chat{\'e}}},\ and\ \bibinfo {author} {\bibfnamefont
  {G.}~\bibnamefont {Theraulaz}},\ }\bibfield  {title} {\bibinfo {title}
  {Deciphering {{Interactions}} in {{Moving Animal Groups}}},\ }\href
  {https://doi.org/10.1371/journal.pcbi.1002678} {\bibfield  {journal}
  {\bibinfo  {journal} {PLoS Computational Biology}\ }\textbf {\bibinfo
  {volume} {8}},\ \bibinfo {pages} {e1002678} (\bibinfo {year}
  {2012})}\BibitemShut {NoStop}%
\bibitem [{\citenamefont {Lord}\ \emph {et~al.}(2016)\citenamefont {Lord},
  \citenamefont {Sun}, \citenamefont {Ouellette},\ and\ \citenamefont
  {Bollt}}]{lord2016InferenceCausalInformation}%
  \BibitemOpen
  \bibfield  {author} {\bibinfo {author} {\bibfnamefont {W.~M.}\ \bibnamefont
  {Lord}}, \bibinfo {author} {\bibfnamefont {J.}~\bibnamefont {Sun}}, \bibinfo
  {author} {\bibfnamefont {N.~T.}\ \bibnamefont {Ouellette}},\ and\ \bibinfo
  {author} {\bibfnamefont {E.~M.}\ \bibnamefont {Bollt}},\ }\bibfield  {title}
  {\bibinfo {title} {Inference of {{Causal Information Flow}} in {{Collective
  Animal Behavior}}},\ }\href {https://doi.org/10.1109/TMBMC.2016.2632099}
  {\bibfield  {journal} {\bibinfo  {journal} {IEEE Transactions on Molecular,
  Biological and Multi-Scale Communications}\ }\textbf {\bibinfo {volume}
  {2}},\ \bibinfo {pages} {107} (\bibinfo {year} {2016})}\BibitemShut {NoStop}%
\bibitem [{\citenamefont {Torney}\ \emph {et~al.}(2018)\citenamefont {Torney},
  \citenamefont {Lamont}, \citenamefont {Debell}, \citenamefont {Angohiatok},
  \citenamefont {Leclerc},\ and\ \citenamefont
  {Berdahl}}]{torney2018InferringRulesSocial}%
  \BibitemOpen
  \bibfield  {author} {\bibinfo {author} {\bibfnamefont {C.~J.}\ \bibnamefont
  {Torney}}, \bibinfo {author} {\bibfnamefont {M.}~\bibnamefont {Lamont}},
  \bibinfo {author} {\bibfnamefont {L.}~\bibnamefont {Debell}}, \bibinfo
  {author} {\bibfnamefont {R.~J.}\ \bibnamefont {Angohiatok}}, \bibinfo
  {author} {\bibfnamefont {L.-M.}\ \bibnamefont {Leclerc}},\ and\ \bibinfo
  {author} {\bibfnamefont {A.~M.}\ \bibnamefont {Berdahl}},\ }\bibfield
  {title} {\bibinfo {title} {Inferring the rules of social interaction in
  migrating caribou},\ }\href {https://doi.org/10.1098/rstb.2017.0385}
  {\bibfield  {journal} {\bibinfo  {journal} {Philosophical Transactions of the
  Royal Society B: Biological Sciences}\ }\textbf {\bibinfo {volume} {373}},\
  \bibinfo {pages} {20170385} (\bibinfo {year} {2018})}\BibitemShut {NoStop}%
\bibitem [{\citenamefont {Lu}\ \emph {et~al.}(2019)\citenamefont {Lu},
  \citenamefont {Zhong}, \citenamefont {Tang},\ and\ \citenamefont
  {Maggioni}}]{lu2019nonparametric}%
  \BibitemOpen
  \bibfield  {author} {\bibinfo {author} {\bibfnamefont {F.}~\bibnamefont
  {Lu}}, \bibinfo {author} {\bibfnamefont {M.}~\bibnamefont {Zhong}}, \bibinfo
  {author} {\bibfnamefont {S.}~\bibnamefont {Tang}},\ and\ \bibinfo {author}
  {\bibfnamefont {M.}~\bibnamefont {Maggioni}},\ }\bibfield  {title} {\bibinfo
  {title} {Nonparametric inference of interaction laws in systems of agents
  from trajectory data},\ }\href {https://doi.org/10.1073/pnas.1822012116}
  {\bibfield  {journal} {\bibinfo  {journal} {Proceedings of the National
  Academy of Sciences}\ }\textbf {\bibinfo {volume} {116}},\ \bibinfo {pages}
  {14424} (\bibinfo {year} {2019})}\BibitemShut {NoStop}%
\bibitem [{\citenamefont {Bhaskar}\ \emph {et~al.}(2019)\citenamefont
  {Bhaskar}, \citenamefont {Manhart}, \citenamefont {Milzman}, \citenamefont
  {Nardini}, \citenamefont {Storey}, \citenamefont {Topaz},\ and\ \citenamefont
  {Ziegelmeier}}]{d2019}%
  \BibitemOpen
  \bibfield  {author} {\bibinfo {author} {\bibfnamefont {D.}~\bibnamefont
  {Bhaskar}}, \bibinfo {author} {\bibfnamefont {A.}~\bibnamefont {Manhart}},
  \bibinfo {author} {\bibfnamefont {J.}~\bibnamefont {Milzman}}, \bibinfo
  {author} {\bibfnamefont {J.~T.}\ \bibnamefont {Nardini}}, \bibinfo {author}
  {\bibfnamefont {K.~M.}\ \bibnamefont {Storey}}, \bibinfo {author}
  {\bibfnamefont {C.~M.}\ \bibnamefont {Topaz}},\ and\ \bibinfo {author}
  {\bibfnamefont {L.}~\bibnamefont {Ziegelmeier}},\ }\bibfield  {title}
  {\bibinfo {title} {Analyzing collective motion with machine learning and
  topology},\ }\href {https://doi.org/10.1063/1.5125493} {\bibfield  {journal}
  {\bibinfo  {journal} {Chaos: An Interdisciplinary Journal of Nonlinear
  Science}\ }\textbf {\bibinfo {volume} {29}},\ \bibinfo {pages} {123125}
  (\bibinfo {year} {2019})}\BibitemShut {NoStop}%
\bibitem [{\citenamefont {Basak}\ \emph {et~al.}(2020)\citenamefont {Basak},
  \citenamefont {Sattari}, \citenamefont {Horikawa},\ and\ \citenamefont
  {Komatsuzaki}}]{basak2020InferringDomainInteractions}%
  \BibitemOpen
  \bibfield  {author} {\bibinfo {author} {\bibfnamefont {U.~S.}\ \bibnamefont
  {Basak}}, \bibinfo {author} {\bibfnamefont {S.}~\bibnamefont {Sattari}},
  \bibinfo {author} {\bibfnamefont {K.}~\bibnamefont {Horikawa}},\ and\
  \bibinfo {author} {\bibfnamefont {T.}~\bibnamefont {Komatsuzaki}},\
  }\bibfield  {title} {\bibinfo {title} {Inferring domain of interactions among
  particles from ensemble of trajectories},\ }\href
  {https://doi.org/10.1103/PhysRevE.102.012404} {\bibfield  {journal} {\bibinfo
   {journal} {Physical Review E}\ }\textbf {\bibinfo {volume} {102}},\ \bibinfo
  {pages} {012404} (\bibinfo {year} {2020})}\BibitemShut {NoStop}%
\bibitem [{\citenamefont {LaChance}\ \emph {et~al.}(2022)\citenamefont
  {LaChance}, \citenamefont {Suh}, \citenamefont {Clausen},\ and\ \citenamefont
  {Cohen}}]{lachance2022LearningRulesCollective}%
  \BibitemOpen
  \bibfield  {author} {\bibinfo {author} {\bibfnamefont {J.}~\bibnamefont
  {LaChance}}, \bibinfo {author} {\bibfnamefont {K.}~\bibnamefont {Suh}},
  \bibinfo {author} {\bibfnamefont {J.}~\bibnamefont {Clausen}},\ and\ \bibinfo
  {author} {\bibfnamefont {D.~J.}\ \bibnamefont {Cohen}},\ }\bibfield  {title}
  {\bibinfo {title} {Learning the rules of collective cell migration using deep
  attention networks},\ }\href {https://doi.org/10.1371/journal.pcbi.1009293}
  {\bibfield  {journal} {\bibinfo  {journal} {PLOS Computational Biology}\
  }\textbf {\bibinfo {volume} {18}},\ \bibinfo {pages} {e1009293} (\bibinfo
  {year} {2022})}\BibitemShut {NoStop}%
\bibitem [{\citenamefont {Nabeel}\ \emph {et~al.}(2023)\citenamefont {Nabeel},
  \citenamefont {Jadhav}, \citenamefont {M}, \citenamefont {Sire},
  \citenamefont {Theraulaz}, \citenamefont {Escobedo}, \citenamefont {Iyer},\
  and\ \citenamefont {Guttal}}]{nabeel2023DatadrivenDiscoveryStochastic}%
  \BibitemOpen
  \bibfield  {author} {\bibinfo {author} {\bibfnamefont {A.}~\bibnamefont
  {Nabeel}}, \bibinfo {author} {\bibfnamefont {V.}~\bibnamefont {Jadhav}},
  \bibinfo {author} {\bibfnamefont {D.~R.}\ \bibnamefont {M}}, \bibinfo
  {author} {\bibfnamefont {C.}~\bibnamefont {Sire}}, \bibinfo {author}
  {\bibfnamefont {G.}~\bibnamefont {Theraulaz}}, \bibinfo {author}
  {\bibfnamefont {R.}~\bibnamefont {Escobedo}}, \bibinfo {author}
  {\bibfnamefont {S.~K.}\ \bibnamefont {Iyer}},\ and\ \bibinfo {author}
  {\bibfnamefont {V.}~\bibnamefont {Guttal}},\ }\bibfield  {title} {\bibinfo
  {title} {Data-driven discovery of stochastic dynamical equations of
  collective motion},\ }\href {https://doi.org/10.1088/1478-3975/ace22d}
  {\bibfield  {journal} {\bibinfo  {journal} {Physical Biology}\ }\textbf
  {\bibinfo {volume} {20}},\ \bibinfo {pages} {056003} (\bibinfo {year}
  {2023})}\BibitemShut {NoStop}%
\bibitem [{\citenamefont {Schaerf}\ \emph {et~al.}(2021)\citenamefont
  {Schaerf}, \citenamefont {{Herbert-Read}},\ and\ \citenamefont
  {Ward}}]{schaerf2021StatisticalMethodIdentifying}%
  \BibitemOpen
  \bibfield  {author} {\bibinfo {author} {\bibfnamefont {T.~M.}\ \bibnamefont
  {Schaerf}}, \bibinfo {author} {\bibfnamefont {J.~E.}\ \bibnamefont
  {{Herbert-Read}}},\ and\ \bibinfo {author} {\bibfnamefont {A.~J.~W.}\
  \bibnamefont {Ward}},\ }\bibfield  {title} {\bibinfo {title} {A statistical
  method for identifying different rules of interaction between individuals in
  moving animal groups},\ }\href {https://doi.org/10.1098/rsif.2020.0925}
  {\bibfield  {journal} {\bibinfo  {journal} {Journal of The Royal Society
  Interface}\ }\textbf {\bibinfo {volume} {18}},\ \bibinfo {pages}
  {rsif.2020.0925, 20200925} (\bibinfo {year} {2021})}\BibitemShut {NoStop}%
\bibitem [{\citenamefont {Lu}\ \emph {et~al.}(2021)\citenamefont {Lu},
  \citenamefont {Maggioni},\ and\ \citenamefont
  {Tang}}]{luLearningInteractionKernels}%
  \BibitemOpen
  \bibfield  {author} {\bibinfo {author} {\bibfnamefont {F.}~\bibnamefont
  {Lu}}, \bibinfo {author} {\bibfnamefont {M.}~\bibnamefont {Maggioni}},\ and\
  \bibinfo {author} {\bibfnamefont {S.}~\bibnamefont {Tang}},\ }\bibfield
  {title} {\bibinfo {title} {Learning interaction kernels in heterogeneous
  systems of agents from multiple trajectories},\ }\href@noop {} {\bibfield
  {journal} {\bibinfo  {journal} {J. Mach. Learn. Res.}\ }\textbf {\bibinfo
  {volume} {22}} (\bibinfo {year} {2021})}\BibitemShut {NoStop}%
\bibitem [{\citenamefont {Messenger}\ \emph {et~al.}(2022)\citenamefont
  {Messenger}, \citenamefont {Wheeler}, \citenamefont {Liu},\ and\
  \citenamefont {Bortz}}]{Messenger2022}%
  \BibitemOpen
  \bibfield  {author} {\bibinfo {author} {\bibfnamefont {D.~A.}\ \bibnamefont
  {Messenger}}, \bibinfo {author} {\bibfnamefont {G.~E.}\ \bibnamefont
  {Wheeler}}, \bibinfo {author} {\bibfnamefont {X.}~\bibnamefont {Liu}},\ and\
  \bibinfo {author} {\bibfnamefont {D.~M.}\ \bibnamefont {Bortz}},\ }\bibfield
  {title} {\bibinfo {title} {Learning anisotropic interaction rules from
  individual trajectories in a heterogeneous cellular population},\ }\href
  {https://doi.org/10.1098/rsif.2022.0412} {\bibfield  {journal} {\bibinfo
  {journal} {Journal of The Royal Society Interface}\ }\textbf {\bibinfo
  {volume} {19}},\ \bibinfo {pages} {20220412} (\bibinfo {year}
  {2022})}\BibitemShut {NoStop}%
\bibitem [{\citenamefont {Nabeel}\ and\ \citenamefont
  {Masila}(2022)}]{nabeel2022DisentanglingIntrinsicMotion}%
  \BibitemOpen
  \bibfield  {author} {\bibinfo {author} {\bibfnamefont {A.}~\bibnamefont
  {Nabeel}}\ and\ \bibinfo {author} {\bibfnamefont {D.~R.}\ \bibnamefont
  {Masila}},\ }\bibfield  {title} {\bibinfo {title} {Disentangling intrinsic
  motion from neighborhood effects in heterogeneous collective motion},\ }\href
  {https://doi.org/10.1063/5.0093682} {\bibfield  {journal} {\bibinfo
  {journal} {Chaos: An Interdisciplinary Journal of Nonlinear Science}\
  }\textbf {\bibinfo {volume} {32}},\ \bibinfo {pages} {063119} (\bibinfo
  {year} {2022})}\BibitemShut {NoStop}%
\bibitem [{\citenamefont {Butail}\ \emph {et~al.}(2016)\citenamefont {Butail},
  \citenamefont {Mwaffo},\ and\ \citenamefont {Porfiri}}]{butail2016model}%
  \BibitemOpen
  \bibfield  {author} {\bibinfo {author} {\bibfnamefont {S.}~\bibnamefont
  {Butail}}, \bibinfo {author} {\bibfnamefont {V.}~\bibnamefont {Mwaffo}},\
  and\ \bibinfo {author} {\bibfnamefont {M.}~\bibnamefont {Porfiri}},\
  }\bibfield  {title} {\bibinfo {title} {Model-free information-theoretic
  approach to infer leadership in pairs of zebrafish},\ }\href
  {https://doi.org/10.1103/PhysRevE.93.042411} {\bibfield  {journal} {\bibinfo
  {journal} {Physical Review E}\ }\textbf {\bibinfo {volume} {93}},\ \bibinfo
  {pages} {042411} (\bibinfo {year} {2016})}\BibitemShut {NoStop}%
\bibitem [{\citenamefont {Mwaffo}\ \emph {et~al.}(2017)\citenamefont {Mwaffo},
  \citenamefont {Butail},\ and\ \citenamefont {Porfiri}}]{mwaffo2017analysis}%
  \BibitemOpen
  \bibfield  {author} {\bibinfo {author} {\bibfnamefont {V.}~\bibnamefont
  {Mwaffo}}, \bibinfo {author} {\bibfnamefont {S.}~\bibnamefont {Butail}},\
  and\ \bibinfo {author} {\bibfnamefont {M.}~\bibnamefont {Porfiri}},\
  }\bibfield  {title} {\bibinfo {title} {Analysis of pairwise interactions in a
  maximum likelihood sense to identify leaders in a group},\ }\href
  {https://doi.org/10.3389/frobt.2017.00035} {\bibfield  {journal} {\bibinfo
  {journal} {Frontiers in Robotics and AI}\ }\textbf {\bibinfo {volume} {4}},\
  \bibinfo {pages} {35} (\bibinfo {year} {2017})}\BibitemShut {NoStop}%
\bibitem [{\citenamefont {Vicsek}\ \emph {et~al.}(1995)\citenamefont {Vicsek},
  \citenamefont {Czir{\'o}k}, \citenamefont {{Ben-Jacob}}, \citenamefont
  {Cohen},\ and\ \citenamefont {Shochet}}]{Vicsek1995}%
  \BibitemOpen
  \bibfield  {author} {\bibinfo {author} {\bibfnamefont {T.}~\bibnamefont
  {Vicsek}}, \bibinfo {author} {\bibfnamefont {A.}~\bibnamefont {Czir{\'o}k}},
  \bibinfo {author} {\bibfnamefont {E.}~\bibnamefont {{Ben-Jacob}}}, \bibinfo
  {author} {\bibfnamefont {I.}~\bibnamefont {Cohen}},\ and\ \bibinfo {author}
  {\bibfnamefont {O.}~\bibnamefont {Shochet}},\ }\bibfield  {title} {\bibinfo
  {title} {Novel type of phase transition in a system of self-driven
  particles},\ }\href {https://doi.org/10.1103/PhysRevLett.75.1226} {\bibfield
  {journal} {\bibinfo  {journal} {Physical Review Letters}\ }\textbf {\bibinfo
  {volume} {75}},\ \bibinfo {pages} {1226} (\bibinfo {year}
  {1995})}\BibitemShut {NoStop}%
\bibitem [{\citenamefont {Ginelli}(2016)}]{physics2016}%
  \BibitemOpen
  \bibfield  {author} {\bibinfo {author} {\bibfnamefont {F.}~\bibnamefont
  {Ginelli}},\ }\bibfield  {title} {\bibinfo {title} {The physics of the
  {{Vicsek}} model},\ }\href {https://doi.org/10.1140/epjst/e2016-60066-8}
  {\bibfield  {journal} {\bibinfo  {journal} {The European Physical Journal
  Special Topics}\ }\textbf {\bibinfo {volume} {225}},\ \bibinfo {pages} {2099}
  (\bibinfo {year} {2016})}\BibitemShut {NoStop}%
\bibitem [{\citenamefont {Czir{\'o}k}\ and\ \citenamefont
  {Vicsek}(2000)}]{vicsek2000}%
  \BibitemOpen
  \bibfield  {author} {\bibinfo {author} {\bibfnamefont {A.}~\bibnamefont
  {Czir{\'o}k}}\ and\ \bibinfo {author} {\bibfnamefont {T.}~\bibnamefont
  {Vicsek}},\ }\bibfield  {title} {\bibinfo {title} {Collective behavior of
  interacting self-propelled particles},\ }\href
  {https://doi.org/10.1016/S0378-4371(00)00013-3} {\bibfield  {journal}
  {\bibinfo  {journal} {Physica A: Statistical Mechanics and its Applications}\
  }\textbf {\bibinfo {volume} {281}},\ \bibinfo {pages} {17} (\bibinfo {year}
  {2000})}\BibitemShut {NoStop}%
\bibitem [{\citenamefont {Chat{\'e}}\ \emph {et~al.}(2008)\citenamefont
  {Chat{\'e}}, \citenamefont {Ginelli}, \citenamefont {Gr{\'e}goire},
  \citenamefont {Peruani},\ and\ \citenamefont {Raynaud}}]{chate2008modeling}%
  \BibitemOpen
  \bibfield  {author} {\bibinfo {author} {\bibfnamefont {H.}~\bibnamefont
  {Chat{\'e}}}, \bibinfo {author} {\bibfnamefont {F.}~\bibnamefont {Ginelli}},
  \bibinfo {author} {\bibfnamefont {G.}~\bibnamefont {Gr{\'e}goire}}, \bibinfo
  {author} {\bibfnamefont {F.}~\bibnamefont {Peruani}},\ and\ \bibinfo {author}
  {\bibfnamefont {F.}~\bibnamefont {Raynaud}},\ }\bibfield  {title} {\bibinfo
  {title} {Modeling collective motion: variations on the vicsek model},\
  }\href@noop {} {\bibfield  {journal} {\bibinfo  {journal} {The European
  Physical Journal B}\ }\textbf {\bibinfo {volume} {64}},\ \bibinfo {pages}
  {451} (\bibinfo {year} {2008})}\BibitemShut {NoStop}%
\bibitem [{\citenamefont {Miguel}\ \emph {et~al.}(2018)\citenamefont {Miguel},
  \citenamefont {Parley},\ and\ \citenamefont
  {Pastor-Satorras}}]{miguel2018effects}%
  \BibitemOpen
  \bibfield  {author} {\bibinfo {author} {\bibfnamefont {M.~C.}\ \bibnamefont
  {Miguel}}, \bibinfo {author} {\bibfnamefont {J.~T.}\ \bibnamefont {Parley}},\
  and\ \bibinfo {author} {\bibfnamefont {R.}~\bibnamefont {Pastor-Satorras}},\
  }\bibfield  {title} {\bibinfo {title} {Effects of heterogeneous social
  interactions on flocking dynamics},\ }\href
  {https://doi.org/10.1103/PhysRevLett.120.068303} {\bibfield  {journal}
  {\bibinfo  {journal} {Physical Review Letters}\ }\textbf {\bibinfo {volume}
  {120}},\ \bibinfo {pages} {068303} (\bibinfo {year} {2018})}\BibitemShut
  {NoStop}%
\bibitem [{\citenamefont {Chatterjee}\ \emph {et~al.}(2023)\citenamefont
  {Chatterjee}, \citenamefont {Mangeat}, \citenamefont {Woo}, \citenamefont
  {Rieger},\ and\ \citenamefont {Noh}}]{chatterjee2023flocking}%
  \BibitemOpen
  \bibfield  {author} {\bibinfo {author} {\bibfnamefont {S.}~\bibnamefont
  {Chatterjee}}, \bibinfo {author} {\bibfnamefont {M.}~\bibnamefont {Mangeat}},
  \bibinfo {author} {\bibfnamefont {C.-U.}\ \bibnamefont {Woo}}, \bibinfo
  {author} {\bibfnamefont {H.}~\bibnamefont {Rieger}},\ and\ \bibinfo {author}
  {\bibfnamefont {J.~D.}\ \bibnamefont {Noh}},\ }\bibfield  {title} {\bibinfo
  {title} {Flocking of two unfriendly species: The two-species vicsek model},\
  }\href {https://doi.org/10.1103/PhysRevE.107.024607} {\bibfield  {journal}
  {\bibinfo  {journal} {Physical Review E}\ }\textbf {\bibinfo {volume}
  {107}},\ \bibinfo {pages} {024607} (\bibinfo {year} {2023})}\BibitemShut
  {NoStop}%
\bibitem [{\citenamefont {D'Orsogna}\ \emph {et~al.}(2006)\citenamefont
  {D'Orsogna}, \citenamefont {Chuang}, \citenamefont {Bertozzi},\ and\
  \citenamefont {Chayes}}]{dorsogna2006SelfPropelledParticlesSoftCore}%
  \BibitemOpen
  \bibfield  {author} {\bibinfo {author} {\bibfnamefont {M.~R.}\ \bibnamefont
  {D'Orsogna}}, \bibinfo {author} {\bibfnamefont {Y.~L.}\ \bibnamefont
  {Chuang}}, \bibinfo {author} {\bibfnamefont {A.~L.}\ \bibnamefont
  {Bertozzi}},\ and\ \bibinfo {author} {\bibfnamefont {L.~S.}\ \bibnamefont
  {Chayes}},\ }\bibfield  {title} {\bibinfo {title} {Self-{{Propelled
  Particles}} with {{Soft-Core Interactions}}: {{Patterns}}, {{Stability}}, and
  {{Collapse}}},\ }\href {https://doi.org/10.1103/PhysRevLett.96.104302}
  {\bibfield  {journal} {\bibinfo  {journal} {Physical Review Letters}\
  }\textbf {\bibinfo {volume} {96}},\ \bibinfo {pages} {104302} (\bibinfo
  {year} {2006})}\BibitemShut {NoStop}%
\bibitem [{\citenamefont {Brown}\ \emph {et~al.}(2017)\citenamefont {Brown},
  \citenamefont {Bossomaier},\ and\ \citenamefont {Barnett}}]{brown2017review}%
  \BibitemOpen
  \bibfield  {author} {\bibinfo {author} {\bibfnamefont {J.~M.}\ \bibnamefont
  {Brown}}, \bibinfo {author} {\bibfnamefont {T.}~\bibnamefont {Bossomaier}},\
  and\ \bibinfo {author} {\bibfnamefont {L.}~\bibnamefont {Barnett}},\
  }\bibfield  {title} {\bibinfo {title} {Review of data structures for
  computationally efficient nearest-neighbour entropy estimators for large
  systems with periodic boundary conditions},\ }\href
  {https://doi.org/10.1016/j.jocs.2017.10.019} {\bibfield  {journal} {\bibinfo
  {journal} {Journal of Computational Science}\ }\textbf {\bibinfo {volume}
  {23}},\ \bibinfo {pages} {109} (\bibinfo {year} {2017})}\BibitemShut
  {NoStop}%
\bibitem [{\citenamefont {Sargsyan}\ \emph {et~al.}(2012)\citenamefont
  {Sargsyan}, \citenamefont {Wright},\ and\ \citenamefont
  {Lim}}]{sargsyan2012GeoPCANewTool}%
  \BibitemOpen
  \bibfield  {author} {\bibinfo {author} {\bibfnamefont {K.}~\bibnamefont
  {Sargsyan}}, \bibinfo {author} {\bibfnamefont {J.}~\bibnamefont {Wright}},\
  and\ \bibinfo {author} {\bibfnamefont {C.}~\bibnamefont {Lim}},\ }\bibfield
  {title} {\bibinfo {title} {{{GeoPCA}}: A new tool for multivariate analysis
  of dihedral angles based on principal component geodesics},\ }\href
  {https://doi.org/10.1093/nar/gkr1069} {\bibfield  {journal} {\bibinfo
  {journal} {Nucleic Acids Research}\ }\textbf {\bibinfo {volume} {40}},\
  \bibinfo {pages} {e25} (\bibinfo {year} {2012})}\BibitemShut {NoStop}%
\bibitem [{\citenamefont {Aghabozorgi}\ \emph {et~al.}(2015)\citenamefont
  {Aghabozorgi}, \citenamefont {Shirkhorshidi},\ and\ \citenamefont
  {Wah}}]{aghabozorgi2015time}%
  \BibitemOpen
  \bibfield  {author} {\bibinfo {author} {\bibfnamefont {S.}~\bibnamefont
  {Aghabozorgi}}, \bibinfo {author} {\bibfnamefont {A.~S.}\ \bibnamefont
  {Shirkhorshidi}},\ and\ \bibinfo {author} {\bibfnamefont {T.~Y.}\
  \bibnamefont {Wah}},\ }\bibfield  {title} {\bibinfo {title} {Time-series
  clustering--a decade review},\ }\href
  {https://doi.org/10.1016/j.is.2015.04.007} {\bibfield  {journal} {\bibinfo
  {journal} {Information systems}\ }\textbf {\bibinfo {volume} {53}},\ \bibinfo
  {pages} {16} (\bibinfo {year} {2015})}\BibitemShut {NoStop}%
\bibitem [{\citenamefont {Jeong}\ \emph {et~al.}(2011)\citenamefont {Jeong},
  \citenamefont {Jeong},\ and\ \citenamefont {Omitaomu}}]{jeong2011weighted}%
  \BibitemOpen
  \bibfield  {author} {\bibinfo {author} {\bibfnamefont {Y.-S.}\ \bibnamefont
  {Jeong}}, \bibinfo {author} {\bibfnamefont {M.~K.}\ \bibnamefont {Jeong}},\
  and\ \bibinfo {author} {\bibfnamefont {O.~A.}\ \bibnamefont {Omitaomu}},\
  }\bibfield  {title} {\bibinfo {title} {Weighted dynamic time warping for time
  series classification},\ }\href
  {https://doi.org/10.1016/j.patcog.2010.09.022} {\bibfield  {journal}
  {\bibinfo  {journal} {Pattern recognition}\ }\textbf {\bibinfo {volume}
  {44}},\ \bibinfo {pages} {2231} (\bibinfo {year} {2011})}\BibitemShut
  {NoStop}%
\bibitem [{\citenamefont {Zhou}\ \emph {et~al.}(2022)\citenamefont {Zhou},
  \citenamefont {Li}, \citenamefont {Li}, \citenamefont {Li},\ and\
  \citenamefont {Li}}]{zhou2022PCAOutperformsPopular}%
  \BibitemOpen
  \bibfield  {author} {\bibinfo {author} {\bibfnamefont {H.~J.}\ \bibnamefont
  {Zhou}}, \bibinfo {author} {\bibfnamefont {L.}~\bibnamefont {Li}}, \bibinfo
  {author} {\bibfnamefont {Y.}~\bibnamefont {Li}}, \bibinfo {author}
  {\bibfnamefont {W.}~\bibnamefont {Li}},\ and\ \bibinfo {author}
  {\bibfnamefont {J.~J.}\ \bibnamefont {Li}},\ }\bibfield  {title} {\bibinfo
  {title} {{{PCA}} outperforms popular hidden variable inference methods for
  molecular {{QTL}} mapping},\ }\href
  {https://doi.org/10.1186/s13059-022-02761-4} {\bibfield  {journal} {\bibinfo
  {journal} {Genome Biology}\ }\textbf {\bibinfo {volume} {23}},\ \bibinfo
  {pages} {210} (\bibinfo {year} {2022})}\BibitemShut {NoStop}%
\bibitem [{\citenamefont {Greenacre}\ \emph {et~al.}(2022)\citenamefont
  {Greenacre}, \citenamefont {Groenen}, \citenamefont {Hastie}, \citenamefont
  {d’Enza}, \citenamefont {Markos},\ and\ \citenamefont
  {Tuzhilina}}]{pca2022}%
  \BibitemOpen
  \bibfield  {author} {\bibinfo {author} {\bibfnamefont {M.}~\bibnamefont
  {Greenacre}}, \bibinfo {author} {\bibfnamefont {P.~J.}\ \bibnamefont
  {Groenen}}, \bibinfo {author} {\bibfnamefont {T.}~\bibnamefont {Hastie}},
  \bibinfo {author} {\bibfnamefont {A.~I.}\ \bibnamefont {d’Enza}}, \bibinfo
  {author} {\bibfnamefont {A.}~\bibnamefont {Markos}},\ and\ \bibinfo {author}
  {\bibfnamefont {E.}~\bibnamefont {Tuzhilina}},\ }\bibfield  {title} {\bibinfo
  {title} {Principal component analysis},\ }\href
  {https://doi.org/10.1038/s43586-022-00184-w} {\bibfield  {journal} {\bibinfo
  {journal} {Nature Reviews Methods Primers}\ }\textbf {\bibinfo {volume}
  {2}},\ \bibinfo {pages} {100} (\bibinfo {year} {2022})}\BibitemShut {NoStop}%
\bibitem [{\citenamefont {Zhang}\ \emph {et~al.}(2017)\citenamefont {Zhang},
  \citenamefont {Li}, \citenamefont {Zong}, \citenamefont {Zhu},\ and\
  \citenamefont {Wang}}]{zhang2017efficient}%
  \BibitemOpen
  \bibfield  {author} {\bibinfo {author} {\bibfnamefont {S.}~\bibnamefont
  {Zhang}}, \bibinfo {author} {\bibfnamefont {X.}~\bibnamefont {Li}}, \bibinfo
  {author} {\bibfnamefont {M.}~\bibnamefont {Zong}}, \bibinfo {author}
  {\bibfnamefont {X.}~\bibnamefont {Zhu}},\ and\ \bibinfo {author}
  {\bibfnamefont {R.}~\bibnamefont {Wang}},\ }\bibfield  {title} {\bibinfo
  {title} {Efficient knn classification with different numbers of nearest
  neighbors},\ }\href {https://doi.org/10.1109/TNNLS.2017.2673241} {\bibfield
  {journal} {\bibinfo  {journal} {IEEE Transactions on Neural Networks and
  Learning Systems}\ }\textbf {\bibinfo {volume} {29}},\ \bibinfo {pages}
  {1774} (\bibinfo {year} {2017})}\BibitemShut {NoStop}%
\bibitem [{\citenamefont {Von~Luxburg}(2007)}]{spec2007}%
  \BibitemOpen
  \bibfield  {author} {\bibinfo {author} {\bibfnamefont {U.}~\bibnamefont
  {Von~Luxburg}},\ }\bibfield  {title} {\bibinfo {title} {A tutorial on
  spectral clustering},\ }\href {https://doi.org/10.1007/s11222-007-9033-z}
  {\bibfield  {journal} {\bibinfo  {journal} {Statistics and Computing}\
  }\textbf {\bibinfo {volume} {17}},\ \bibinfo {pages} {395} (\bibinfo {year}
  {2007})}\BibitemShut {NoStop}%
\bibitem [{\citenamefont {Rousseeuw}(1987)}]{sil1987}%
  \BibitemOpen
  \bibfield  {author} {\bibinfo {author} {\bibfnamefont {P.~J.}\ \bibnamefont
  {Rousseeuw}},\ }\bibfield  {title} {\bibinfo {title} {Silhouettes: {{A}}
  graphical aid to the interpretation and validation of cluster analysis},\
  }\href {https://doi.org/10.1016/0377-0427(87)90125-7} {\bibfield  {journal}
  {\bibinfo  {journal} {Journal of Computational and Applied Mathematics}\
  }\textbf {\bibinfo {volume} {20}},\ \bibinfo {pages} {53} (\bibinfo {year}
  {1987})}\BibitemShut {NoStop}%
\bibitem [{\citenamefont {Orange}\ and\ \citenamefont
  {Abaid}(2015)}]{orange2015transfer}%
  \BibitemOpen
  \bibfield  {author} {\bibinfo {author} {\bibfnamefont {N.}~\bibnamefont
  {Orange}}\ and\ \bibinfo {author} {\bibfnamefont {N.}~\bibnamefont {Abaid}},\
  }\bibfield  {title} {\bibinfo {title} {A transfer entropy analysis of
  leader-follower interactions in flying bats},\ }\href
  {https://doi.org/10.1140/epjst/e2015-50235-9} {\bibfield  {journal} {\bibinfo
   {journal} {The European Physical Journal Special Topics}\ }\textbf {\bibinfo
  {volume} {224}},\ \bibinfo {pages} {3279} (\bibinfo {year}
  {2015})}\BibitemShut {NoStop}%
\bibitem [{\citenamefont {Fujii}\ \emph {et~al.}(2021)\citenamefont {Fujii},
  \citenamefont {Takeishi}, \citenamefont {Tsutsui}, \citenamefont {Fujioka},
  \citenamefont {Nishiumi}, \citenamefont {Tanaka}, \citenamefont {Fukushiro},
  \citenamefont {Ide}, \citenamefont {Kohno}, \citenamefont {Yoda},
  \citenamefont {Takahashi}, \citenamefont {Hiryu},\ and\ \citenamefont
  {Kawahara}}]{fujii2021learning}%
  \BibitemOpen
  \bibfield  {author} {\bibinfo {author} {\bibfnamefont {K.}~\bibnamefont
  {Fujii}}, \bibinfo {author} {\bibfnamefont {N.}~\bibnamefont {Takeishi}},
  \bibinfo {author} {\bibfnamefont {K.}~\bibnamefont {Tsutsui}}, \bibinfo
  {author} {\bibfnamefont {E.}~\bibnamefont {Fujioka}}, \bibinfo {author}
  {\bibfnamefont {N.}~\bibnamefont {Nishiumi}}, \bibinfo {author}
  {\bibfnamefont {R.}~\bibnamefont {Tanaka}}, \bibinfo {author} {\bibfnamefont
  {M.}~\bibnamefont {Fukushiro}}, \bibinfo {author} {\bibfnamefont
  {K.}~\bibnamefont {Ide}}, \bibinfo {author} {\bibfnamefont {H.}~\bibnamefont
  {Kohno}}, \bibinfo {author} {\bibfnamefont {K.}~\bibnamefont {Yoda}},
  \bibinfo {author} {\bibfnamefont {S.}~\bibnamefont {Takahashi}}, \bibinfo
  {author} {\bibfnamefont {S.}~\bibnamefont {Hiryu}},\ and\ \bibinfo {author}
  {\bibfnamefont {Y.}~\bibnamefont {Kawahara}},\ }\bibfield  {title} {\bibinfo
  {title} {Learning interaction rules from multi-animal trajectories via
  augmented behavioral models},\ }in\ \href
  {https://proceedings.neurips.cc/paper_files/paper/2021/file/5c572eca050594c7bc3c36e7e8ab9550-Paper.pdf}
  {\emph {\bibinfo {booktitle} {Advances in Neural Information Processing
  Systems}}},\ Vol.~\bibinfo {volume} {34},\ \bibinfo {editor} {edited by\
  \bibinfo {editor} {\bibfnamefont {M.}~\bibnamefont {Ranzato}}, \bibinfo
  {editor} {\bibfnamefont {A.}~\bibnamefont {Beygelzimer}}, \bibinfo {editor}
  {\bibfnamefont {Y.}~\bibnamefont {Dauphin}}, \bibinfo {editor} {\bibfnamefont
  {P.}~\bibnamefont {Liang}},\ and\ \bibinfo {editor} {\bibfnamefont {J.~W.}\
  \bibnamefont {Vaughan}}}\ (\bibinfo  {publisher} {Curran Associates, Inc.},\
  \bibinfo {year} {2021})\ pp.\ \bibinfo {pages} {11108--11122}\BibitemShut
  {NoStop}%
\bibitem [{\citenamefont {Wang}\ \emph {et~al.}(2016)\citenamefont {Wang},
  \citenamefont {Yao},\ and\ \citenamefont {Zhao}}]{wang2016auto}%
  \BibitemOpen
  \bibfield  {author} {\bibinfo {author} {\bibfnamefont {Y.}~\bibnamefont
  {Wang}}, \bibinfo {author} {\bibfnamefont {H.}~\bibnamefont {Yao}},\ and\
  \bibinfo {author} {\bibfnamefont {S.}~\bibnamefont {Zhao}},\ }\bibfield
  {title} {\bibinfo {title} {Auto-encoder based dimensionality reduction},\
  }\href {https://doi.org/10.1016/j.neucom.2015.08.104} {\bibfield  {journal}
  {\bibinfo  {journal} {Neurocomputing}\ }\textbf {\bibinfo {volume} {184}},\
  \bibinfo {pages} {232} (\bibinfo {year} {2016})}\BibitemShut {NoStop}%
\bibitem [{\citenamefont {Yu}\ \emph {et~al.}(2019)\citenamefont {Yu},
  \citenamefont {Si}, \citenamefont {Hu},\ and\ \citenamefont
  {Zhang}}]{yu2019review}%
  \BibitemOpen
  \bibfield  {author} {\bibinfo {author} {\bibfnamefont {Y.}~\bibnamefont
  {Yu}}, \bibinfo {author} {\bibfnamefont {X.}~\bibnamefont {Si}}, \bibinfo
  {author} {\bibfnamefont {C.}~\bibnamefont {Hu}},\ and\ \bibinfo {author}
  {\bibfnamefont {J.}~\bibnamefont {Zhang}},\ }\bibfield  {title} {\bibinfo
  {title} {A review of recurrent neural networks: Lstm cells and network
  architectures},\ }\href {https://doi.org/10.1162/neco_a_01199} {\bibfield
  {journal} {\bibinfo  {journal} {Neural Computation}\ }\textbf {\bibinfo
  {volume} {31}},\ \bibinfo {pages} {1235} (\bibinfo {year}
  {2019})}\BibitemShut {NoStop}%
\bibitem [{\citenamefont {Toner}\ and\ \citenamefont
  {Tu}(1998)}]{toner1998flocks}%
  \BibitemOpen
  \bibfield  {author} {\bibinfo {author} {\bibfnamefont {J.}~\bibnamefont
  {Toner}}\ and\ \bibinfo {author} {\bibfnamefont {Y.}~\bibnamefont {Tu}},\
  }\bibfield  {title} {\bibinfo {title} {Flocks, herds, and schools: A
  quantitative theory of flocking},\ }\href
  {https://doi.org/10.1103/PhysRevE.58.4828} {\bibfield  {journal} {\bibinfo
  {journal} {Physical Review E}\ }\textbf {\bibinfo {volume} {58}},\ \bibinfo
  {pages} {4828} (\bibinfo {year} {1998})}\BibitemShut {NoStop}%
\bibitem [{\citenamefont {Zagli}\ \emph {et~al.}(2023)\citenamefont {Zagli},
  \citenamefont {Pavliotis}, \citenamefont {Lucarini},\ and\ \citenamefont
  {Alecio}}]{zagli2023DimensionReductionNoisy}%
  \BibitemOpen
  \bibfield  {author} {\bibinfo {author} {\bibfnamefont {N.}~\bibnamefont
  {Zagli}}, \bibinfo {author} {\bibfnamefont {G.~A.}\ \bibnamefont
  {Pavliotis}}, \bibinfo {author} {\bibfnamefont {V.}~\bibnamefont
  {Lucarini}},\ and\ \bibinfo {author} {\bibfnamefont {A.}~\bibnamefont
  {Alecio}},\ }\bibfield  {title} {\bibinfo {title} {Dimension reduction of
  noisy interacting systems},\ }\href
  {https://doi.org/10.1103/PhysRevResearch.5.013078} {\bibfield  {journal}
  {\bibinfo  {journal} {Physical Review Research}\ }\textbf {\bibinfo {volume}
  {5}},\ \bibinfo {pages} {013078} (\bibinfo {year} {2023})}\BibitemShut
  {NoStop}%
\end{thebibliography}%

\end{document}